\documentclass[prb,twocolumn,showpacs,amsmath,amssymb]{revtex4}
\usepackage{graphics}
\usepackage{epsfig}

\begin{document}

\title{The even-odd effect in short antiferromagnetic Heisenberg chains}

\author{A. Machens}

\author{N. P. Konstantinidis}
\altaffiliation{Present address: Fachbereich Physik und Landesforschungszentrum OPTIMAS, Technische Universit\"at
Kaiserslautern, 67663 Kaiserslautern, Germany}

\author{O. Waldmann}
\email{oliver.waldmann@physik.uni-freiburg.de}
\affiliation{Physikalisches Institut, Universit\"at Freiburg, 79104
Freiburg, Germany}

\author{I. Schneider}
\email{imke.schneider@tu-dresden.de}
\affiliation{Institut f\"ur Theoretische Physik, Technische Universit\"at Dresden, 01062 Dresden, Germany}

\author{S. Eggert}
\affiliation{Fachbereich Physik und Landesforschungszentrum OPTIMAS,
Technische Universit\"at Kaiserslautern, 67663 Kaiserslautern, Germany}

\date{\today}
%\date{20.08.2012}

\begin{abstract}
Motivated by recent experiments on chemically synthesized magnetic molecular chains we investigate the lowest lying
energy band of short spin-$s$ antiferromagnetic Heisenberg chains focusing on effects of open boundaries. By numerical
diagonalization we find that the Land\'e pattern in the energy levels, i.e. $E(S)\propto S(S+1)$ for total spin $S$,
known from e.g. ring-shaped nanomagnets, can be recovered in odd-membered chains while strong deviations are found for
the lowest excitations in chains with an even number of sites. This particular even-odd effect in the short Heisenberg
chains cannot be explained by simple effective Hamiltonians and symmetry arguments. We go beyond these approaches,
taking into account quantum fluctuations by means of a path integral description and the valence bond basis, but the
resulting quantum edge-spin picture which is known to work well for long chains does not agree with the numerical
results for short chains and cannot explain the even-odd effect. Instead, by analyzing also the classical chain model,
we show that spatial fluctuations dominate the physical behavior in short chains, with length $N \alt e^{\pi s}$, for
any spin $s$. Such short chains are found to display a unique behavior, which is not related to the thermodynamic limit
and cannot be described well by theories developed for this regime.
\end{abstract}


%# 75.50.Xx
%Molecular magnets
%# 75.10.Jm
%Quantized spin models (magnetism)
%# 78.70.Nx
%Neutron inelastic scattering (condensed matter)
\maketitle

\section{\label{sec:intro} Introduction}

The antiferromagnetic Heisenberg model is the appropriate starting point to understand magnetism in a large variety
of different materials where strong electron correlations are important. Within its context the quantum nature of magnetism
reveals itself, leading to a vast array of fascinating phenomena. Accordingly this model has been the topic of numerous
experimental and theoretical works in different combinations of magnetic lattices and spin magnitudes. Indeed, many
interesting theoretical concepts have been developed specifically for the antiferromagnetic Heisenberg model in the
thermodynamic limit,\cite{Manousakis91,Auerbach98,Lhuillier02}  including spin waves, effective Hamiltonians, quantum
field theories, and hydrodynamic methods. However, as will be shown in this paper most of these methods fail when
describing finite (but possibly large) clusters of spins, which show a unique behavior that strongly depends on
the boundary conditions and topology and therefore form a separate class of magnetic materials.

The properties of the antiferromagnetic Heisenberg model on small clusters such as dimers, trimers and tetramers, which
are relatively simple to understand, were mainly of relevance to chemists, as the Heisenberg exchange describes
magnetic interactions between metal ions in polynuclear metal complexes very well,\cite{Bencini90} and also allowed the
investigation of the origin of the fundamental magnetic interactions.\cite{Wal12-rmp} However, the advances in the
synthesis of metal complexes during the last fifteen years have produced nanosized magnetic molecules with up to a few
dozens of magnetic metal ions interacting with each other via Heisenberg exchange, creating the new class of molecular
nanomagnets.\cite{DanteGatteschiSci,WaldmannRev} Molecular nanomagnets belong to the mesoscopic regime, and their
magnetism can be considerably more complex than in few-membered metal clusters and possess a quantum many-body
character, yet they may not contain enough spin centers to be appropriately described by the methods and techniques
developed for extended systems. Apart from synthetic chemistry, artificial engineering of quantum spin clusters has
also emerged in recent years.\cite{Hirjibehedin06,Trimersurface,sufacerev} Here clusters of magnetic ions have been
fabricated directly on insulating surfaces, and their magnetic properties were measured with scanning tunneling
microscopy.

Nanosized spin clusters are ideal to study basic questions of quantum mechanics in mesoscopic systems, such as the
efficiency of different metal centers or topologies towards a desired magnetic property, or the transition from the
quantum to the classical regime for larger ion spins.\cite{Review10,Wal12-rmp} They also provide an ideal testing
ground for the validity of different theoretical models. In this paper we consider the antiferromagnetic Heisenberg
model with a small number of spins of magnitude $s$ placed along a chain of length $N$ with open ends,
\begin{equation}
\label{eq:Hchain}
H = J \sum_{i=1}^{N-1}{ \mathbf{s}_i \cdot \mathbf{s}_{i+1}},
\end{equation}
where $\mathbf{s}_i$ denotes a spin-$s$ operator on site $i$ and $J > 0$ for antiferromagnetic interactions. We will
show that despite the apparent simplicity of the model the detailed aspects of the spectrum are highly non-trivial.

The model for an infinite chain or one-dimensional antiferromagnetic Heisenberg chain (AFHC) has attracted enormous
interest, especially after Haldane's conjecture of a fundamental difference between integer and half-integer spin
chains, which is by now well established.\cite{Haldane,Haldane2,AffRev} Finite but long chains have also received
significant attention, but mainly for connecting numerical results to the thermodynamic limit via scaling
\cite{QinNgSu} or for understanding boundary,\cite{edgesusc} impurity,\cite{anfuso2003} and doping
effects.\cite{PhysRevLett.89.047202} On the other hand, very little has been done for short chains with $s > 1/2$, in
which the Haldane gap is always smaller than the finite-size excitation energy, and the difference between integer and
half-integer spins is thus expected to become irrelevant.

Short AFHCs have recently become accessible experimentally, e.g., through the synthesis of molecular Cr$^{3+}$
($s=3/2$) modified wheels and
horseshoes.\cite{Cr6,Cr7,Ghi07-chains,Bia09-cr8zn,Fur08-cr8cr8cd,Pil07-cr7cdhfepr,Mic06-cr7cdnmr} Two Fe$^{3+}$
containing ring molecules were also studied, which magnetically represent $s=5/2$ chains of lengths $N= 7$ and
$17$.\cite{Gui07-fe7zn,Hen08-fe17ga} Magnetic and inelastic neutron scattering measurements on the Cr$_6$ and Cr$_7$
horseshoes ($s = 3/2$, $N=6$ and $7$) have provided a detailed view on the magnetic excitation spectrum of these
clusters,\cite{Cr6,Cr7} and pointed to systematic differences between chains with even and odd number of spins. The
expected even-odd effects were found, such as total spin $S = 0$ and $S = s$ in the ground state for even and odd $N$
respectively, which can obviously be associated to the interplay of the open boundary conditions and the symmetry
requirements. However, a more subtle but striking difference in the energy spectrum has also been noted in addition,
which we here refer to as the even-odd effect.

This even-odd effect manifests itself in the lowest energy band as a function of total spin $S$, which depending on the
context is known as the tower of states, quasi-degenerate joint states, or the $L$
band.\cite{And52,Bernu92,RotationalBands,HabRing} In this band the energies are expected to increase as $E(S) \propto
S(S+1)$, and for bipartite spin systems it is generally possible to use a simplified model, the $H_{AB}$ Hamiltonian,
to describe it. The $H_{AB}$ model has been very successfully applied to even-membered antiferromagnetic Heisenberg
rings (odd-membered rings will not be considered in this work),\cite{HabRing} and experimentally confirmed in fine
detail for the Cr$_8$, CsFe$_8$ and Fe$_{18}$ molecular wheels.\cite{Cr8,CsFe8,Fe18} Further examples are the
Mn-[3$\times$3] grid and the Fe$_{30}$ Keplerate
molecules.\cite{Mn3x3,Sch01-fe30rbmodel,Exl03-fe30dmrg,Gar06-fe30ins,OWFe30} For short chains with open ends, however,
the $E(S) \propto S(S+1)$ approximation of the $L$ band appears to work very well for \emph{odd} chains, while for
\emph{even} chains it surprisingly fails in the low energy sector.

We here explore alternative approaches to extract the essential physics of the AFHC model for short chains. In
particular, we carefully reanalyze the $H_{AB}$ model, and compare numerical diagonalization results for
Eq.~\eqref{eq:Hchain} with the predictions of field theoretical approaches, the classical AFHC, and parent valence-bond
Hamiltonians. These approaches differ in their treatment of the quantum and spatial fluctuations and allow us to
get insight into the roles played by them.

As main results we will demonstrate that the edge-spin picture, which has been firmly established for long chains
for $s=1$ \cite{kennedy1990,hagiwara1990,glarum1991,Sorensen1994} and higher $s$,\cite{NgNLSM,QinNgSu,LouQinNgSu} and
was conjectured to stay robust up to very small $N$,\cite{QinNgSu,LouQinNgSu} is not consistent with our exact
numerical data, justifying our distinction into short and long chains. Furthermore, for spins $s \ge 3/2$ the even-odd
effect has converged already closely to the classical behavior, yet quantum fluctuations are still noticeable. We
finally conclude that short AFHCs posses simultaneously a classical and a quantum character, blurring the limit between
classical and quantum behavior.

The paper is organized as follows: in Sec.~\ref{sec:evenodd} the even-odd effect is introduced, and the $H_{AB}$ model
is analyzed in Sec.~\ref{sec:AB}. Section~\ref{sec:nlsm} deals with the predictions of the O(3) nonlinear
$\sigma$-model, which originates in a path integral representation. Section~\ref{sec:spinhalf} goes away from
semiclassical descriptions to investigate quantum mechanical effects on the extreme quantum $s=1/2$ case. The classical
limit of the AFHC is investigated in Sec.~\ref{sec:classicallimit}, and spin density and correlation effects in
Sec.~\ref{sec:spindensity}. The valence-bond model description of the AFHC is analyzed in Sec.~\ref{sec:vbs} and
finally, Sec.~\ref{sec:concl} presents the conclusions. The Appendices that then follow include relevant information on
various topics of the main text.

%%%%%%%%%%%%%%%%%%%%%%%%%%%%%%%%%%%%%%%%%%%%%%%%%%%%%%%%%%%%%%%%%%
% Even-Odd Effect
%%%%%%%%%%%%%%%%%%%%%%%%%%%%%%%%%%%%%%%%%%%%%%%%%%%%%%%%%%%%%%%%%%
\section{\label{sec:evenodd} The Even-Odd Effect}

\begin{figure*}
\includegraphics[width=14cm]{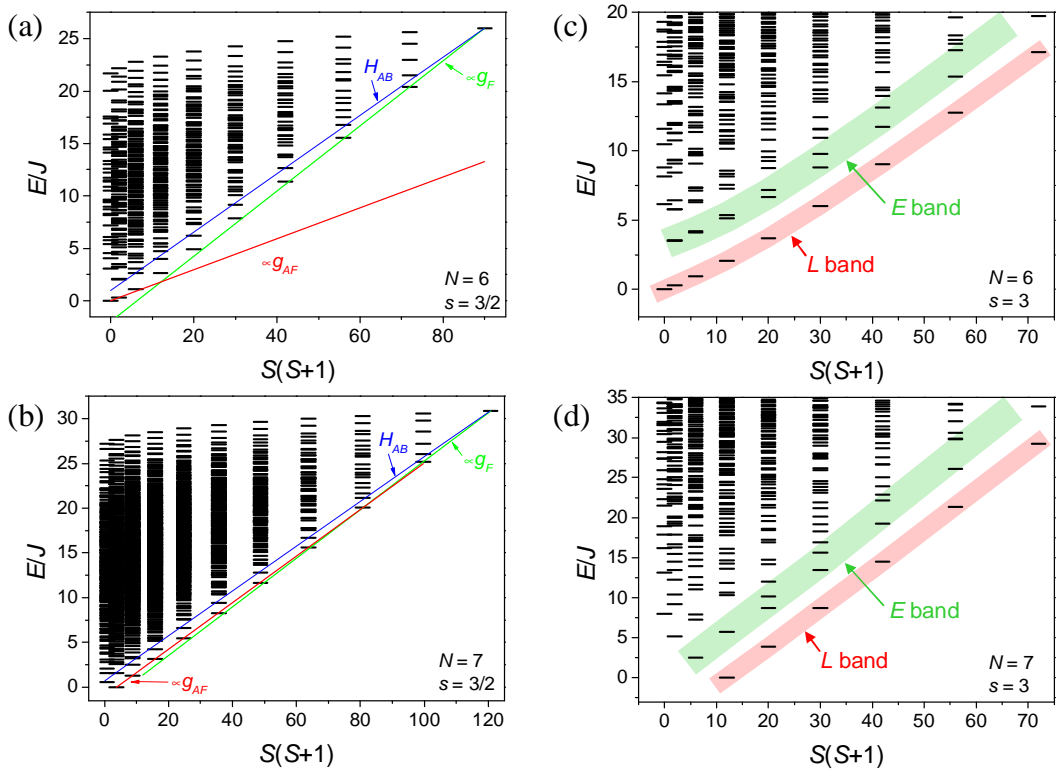}
\caption{\label{fig:spectra} (Color online)
Energy spectra of antiferromagnetic Heisenberg chains as function of
$S(S+1)$ for (a) $N = 6$, $s=3/2$, (b) $N = 7$, $s=3/2$, (c) $N = 6$, $s=3$, and (d) $N = 7$, $s=3$. Panels (a) and (b)
show the full spectrum, while (c) and (d) the low lying spectrum. In panels (a) and (b), the solid lines indicate the
slopes $g_{AF}$ (red line), $g_{F}$ (green line), and the prediction of the $H_{AB}$ model (blue line), which
establishes also an upper bound for the minimal states in each $S$ sector. The $L$ band is clearly visible in all
panels; in panels (c) and (d) it is emphasized together with the $E$ band by the thick underlying bars. }
\end{figure*}

The different symmetry properties between even- and odd-membered AFHCs naturally generate differences in the magnetic
properties, which show up in various quantities giving rise to various "even-odd" effects. The most obvious difference
is in the ground state, which has total spin $S_g = 0$ for even and $S_g = s$ for odd chains due to the residual spin,
as can be inferred from the classical antiferromagnetic spin configuration in the ground state, or the theorem of Lieb
and Mattis.\cite{LiebMattis} Pronounced differences are also present for example in the spin density and spin
correlation functions (see Sec.~\ref{sec:spindensity}), which are however largely dictated by symmetry and boundary
considerations.

The even-odd effect observed in this paper is related to pronounced deviations from an approximate $E(S) \propto
S(S+1)$ energy dependence, and through this is associated to the $H_{AB}$ model, or more generally the $L$- and
$E$-band picture of the excitations in not too large bipartite spin clusters (the $E$-band collectively denotes the
next higher lying rotational bands).\cite{WaldmannRev} In this section we define and characterize the even-odd effect.
The $H_{AB}$ model and the $L\&E$-band picture will be described and reanalyzed in more detail in the following
section.

The Hamiltonian of Eq.~\eqref{eq:Hchain} is SU(2) invariant and the eigenstates are organized as multiplets of $2S+1$
degenerate states according to $S$. We focus on the lowest spin multiplet in each $S$ sector, i.e., the $L$ band.
Starting from the ground state, the energies of the lowest spin multiplets increase approximately as $S(S+1)$
(consistent with the Lieb-Mattis ordering of energies \cite{LiebMattis}). They thus fall approximately on a straight
line in an energy vs. $S(S+1)$ plot, which is the $L$ band. The ground state spin will be denoted by $S_g$, and that of
the state with maximal spin or ferromagnetic state by $S_f = Ns$. The region in the energy spectrum with small (large)
values of $S$ will be called the antiferromagnetic (ferromagnetic) region. Energy values are always presented with
respect to the ground-state energy. The eigenvalues and eigenstates of Eq.~\eqref{eq:Hchain} have been calculated by
full exact, subspace iteration and/or Lanczos diagonalization, with the Hamiltonian matrix represented in the product
space or by irreducible tensor operators \cite{Bencini90}, and spatial symmetry was also used \cite{Wal00-sym,NPK05}
(see Appendix~\ref{sec:base} for more details).

For illustration, Fig.~\ref{fig:spectra} shows spectra of AFHCs with $N=6$ and $7$ for both $s=3/2$ and $3$. While it
is clearly possible to identify the $L$ band, there are systematic deviations. In particular, for $N=6$
the $L$ band is no longer well approximated by an $S(S+1)$ dependence for small $S$. In contrast, for $N=7$ the
$S(S+1)$ behavior is surprisingly well obeyed. Depending on the view point, the question hence arises why the $L$ band
deviates from the $S(S+1)$ behavior so strongly in the even chain, or why is it so well realized in the odd chain.

In order to characterize the differences between even and odd $N$ we consider the (normalized) slope of the $L$ band in
the $E$ vs. $S(S+1)$ representation as a function of $S$,
\begin{eqnarray}
\label{def:normalizedslope}
  g(S)= \frac{2}{\Delta_{AB}} \frac{ E(S)-E(S-1) }{2S}
\end{eqnarray}
with $S \ge S_g$, which is the discretized version of $g(S) = \frac{ 2 }{ \Delta_{AB} } \partial E /
\partial[S(S+1)]$. The slope is normalized with the quantity $\Delta_{AB}/2$, which is given for even and odd chains, and
rings (for comparison), as
\begin{eqnarray}
\label{eq:DABdef}
  \text{even chain: } &\Delta_{AB}& =\frac{4 J(N-1)}{N^2},\nonumber\\
  \text{odd chain: }  &\Delta_{AB}& =\frac{4J}{N+1},\nonumber\\
  \text{even ring: } &\Delta_{AB}& =\frac{4 J}{N}.
\end{eqnarray}
The normalization allows us to directly compare with the predictions of the $H_{AB}$ model, for which $g(S)=1$ (see
Sec.~\ref{sec:AB}). It is noted that the difference of $\Delta_{AB}$ for the even ring and the chains is of order
$O(1/N^2)$ while that for the even and odd chain is of order $O(1/N^3)$, and that the slope emphasizes differences in
the antiferromagnetic region.

Experimentally, the energies $E(S)$ are directly connected to the low-temperature magnetization curve $M(B)$, and their
differences can be extracted from it ($B$ is the magnetic field). In a magnetic field the energies are modified as
$E_B(S,B) = E(S) - 2 \mu_B B S$ (assuming a gyromagnetic factor equal to 2), and at zero temperature $M$ is
discontinuous with steps of height $2 \mu_B$ at fields $B_S = [E(S)-E(S-1)]/(2 \mu_B)$. An energy dependence $E(S) =
\frac{1}{2}\Delta S(S+1)$ with an appropriate gap $\Delta$ yields steps at regularly spaced fields $B_S =
\frac{\Delta}{2 \mu_B} S$, and deviations from the $S(S+1)$ energy dependence or a constant slope $g(S)$ are detected
as deviations from this regular field pattern. If one extrapolates through the magnetization steps or measures the
magnetization at elevated temperatures, the $M(B)$ curve increases linearly with field (except at very low fields), and
deviations from the $E(S) \propto S(S+1)$ behavior are observed as a field-dependent slope or non-constant
susceptibility $M/B$. For the classical case $s \rightarrow \infty$ the slope can in fact rigorously be shown to be
proportional to the inverse susceptibility (Sec.~\ref{sec:classicallimit}). The energy differences $E(S)-E(S-1)$ can
also be measured directly by inelastic neutron scattering due to the selection rule $|S-S'| = 1$, permitting a
spectroscopic determination of the slope $g(S)$.

\begin{figure}
\includegraphics[width=7cm]{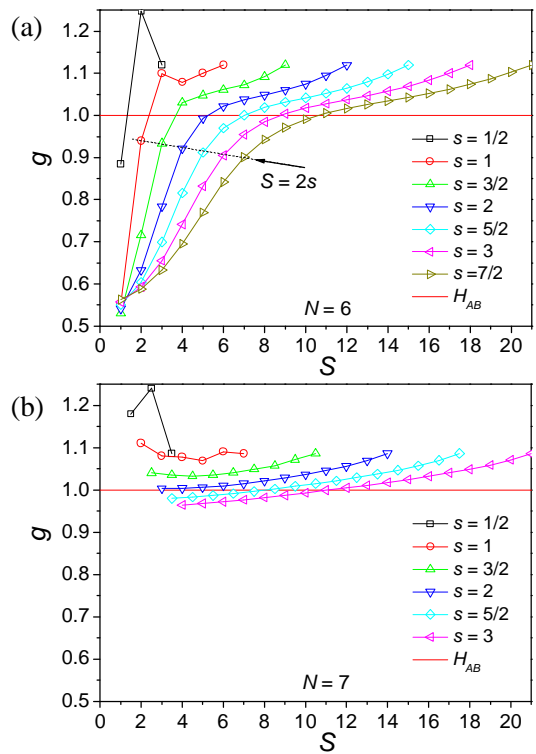}
\caption{\label{fig:slopesevenodd}(Color online)
Slope $g(S)$ as function of $S$ for $S \ge S_g$ for (a) $N=6$ and (b) $N=7$ chains for different spins $s$
(symbols). The horizontal (red) line is the slope $g(S)=1$ of the $H_{AB}$ model. The arrow and the dotted line in (a)
point to the value of the slope for $S=2s$. Lines are guides to the eye.}
\end{figure}

The slope $g(S)$ is plotted for $N=6$ and $7$ for $s$ ranging from 1/2 to 7/2 or 3 in Fig.~\ref{fig:slopesevenodd}. For
$N=7$, which represents odd chains, $g(S)$ varies weakly with $S$ (at least for $s \ge 3/2$), and is close to 1. In
contrast, there is a strong reduction of the slope for $N=6$, the representative for even chains, for $S \lesssim 2s$.
In comparison to the maximal total spin $S_f$ the region of strong deviation is thus given by $S/S_f \lesssim 2/N$, and
the disagreement with the $S(S+1)$ spectrum does not alleviate with increasing $s$, contrary to naive expectation. The
deviation is pronounced for a wide range of $S$ values for short chains, but for long chains where $s \ll S_f$ it
becomes very limited in range and is less important. These observations are a major result of this paper and establish
the even-odd effect considered here.

\begin{figure}
\includegraphics[width=7cm]{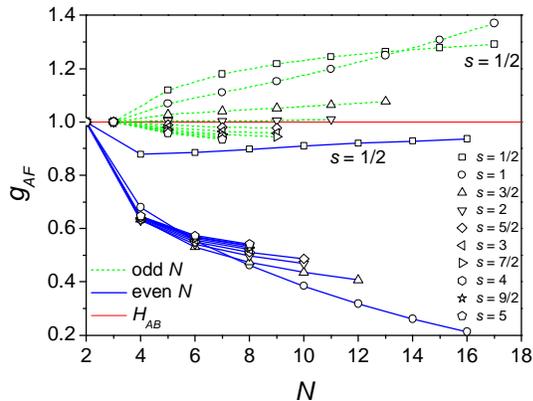}
\caption{\label{fig:gAF} (Color online)
Slope $g_{AF}$ for even (blue lines) and odd (green dashed lines) chains as a function of the length $N$ for different
spins $s$ (symbols). The horizontal (red) line represents the slope $g(S)=1$ predicted from the $H_{AB}$ model. Lines are guides
to the eye.}
\end{figure}

The even-odd effect is most pronounced in the antiferromagnetic region and hence also the slopes
\begin{eqnarray}
\label{eq:gAFgF}
  g_{AF}&=& g( S_{g}+1 ), \nonumber\\
  g_{F}&=& g( S_{f} )
\end{eqnarray}
are considered, which characterize the antiferromagnetic and ferromagnetic parts of the $L$ band respectively [see
Figs~\ref{fig:spectra}(a), (b)]. In Fig.~\ref{fig:gAF} $g_{AF}$ is plotted as function of $N$ for various $s$. Since
$\Delta_{AB}$ in Eq.~(\ref{eq:DABdef}) is slightly smaller for even than for odd chains, the scaling by $\Delta_{AB}$
mitigates the even-odd difference in $g_{AF}$ as compared to the non-normalized slope. Nevertheless, the difference
between even and odd chains is obvious. Interestingly, the difference between the even and odd $g_{AF}$ grows with $N$. This
divergence suggests that the excitations on the $L$ band in the antiferromagnetic region have different physical origin
for even and odd chains. For much larger $N$ the behavior known for long chains will be approached. The
ferromagnetic slope $g_F$ can be calculated exactly from the lowest one-magnon energy as\cite{AffRev}
\begin{equation}
\label{eq:gFexact}
  g_F = \frac{2 J}{N \Delta_{AB}} \left[ 1- \cos\left( \pi \frac{N-1}{N} \right) \right].
\end{equation}
Hence, as $g_F$ is just the classical one-magnon energy for a particular wave vector, the even-odd effect is absent in
the ferromagnetic regime. We note that using Eq.~(\ref{eq:gFexact}) one finds $g_F > 1$ for $N > 3$. This is an
important difference to even-membered rings, where due to translational invariance the eigenvalues and eigenfunctions
of the lowest one-magnon state are equivalent to those of $H_{AB}$ and hence $g_F^{ring}=1$ for all ring
sizes.\cite{HabRing} The absence of the even-odd effect in the ferromagnetic region is also seen in the spin density in
the lowest one-magnon state, given by
\begin{equation}
 \label{eq:spindensityFexact}
  \langle s_i^z \rangle = s - \frac{2}{N}\cos\left(\pi \frac{2i-1}{2N}\right)^2.
\end{equation}
There is no qualitative difference between even and odd $N$ here.

\begin{figure}
\includegraphics[width=7cm]{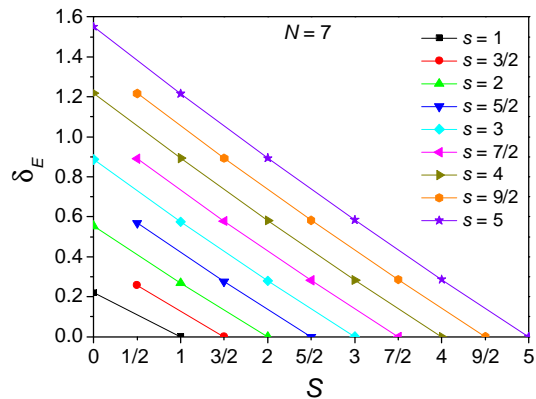}
\caption{\label{fig:edgeodd}(Color online)
Normalized excitation energies $\delta_E(s)$ of the lowest energy states in the spin sectors $S \leq S_g$ for the $N=7$
chain and spins $s$ ranging from 1 to 5 (symbols). For each chain an approximate linear dependence on $S$ is observed. Lines are
guides to the eye.}
\end{figure}

It is also of interest to consider the lowest energies $E(S)$ in odd chains for $S \leq S_g = s$, since the $H_{AB}$
model (Sec. \ref{sec:AB}) and the edge-spin picture (Sec. \ref{sec:nlsm}) predict distinctly different trends with $S$
for these states. The energies in this region can be represented by the normalized excitation energies
\begin{equation}
\label{eq:DeltaEodd}
  \delta_{E}(S) = \frac{2 \left[ E(S) - E(S_g) \right]}{ \Delta_{AB} (N-1) s},
\end{equation}
shown in Fig.~\ref{fig:edgeodd} for the $N=7$ chain for $s=1$ to 5. The energies get smaller with increasing $S$,
$E(S)>E(S+1)$, and the dependence on $S$ is essentially linear with a small curvature.

We conclude this section by discussing the different symmetry properties of even and odd chains. The mirror symmetry
about a central site for odd chains can support a N\'eel-type antiferromagnetic configuration in the ground state [as
indicated in Fig.~\ref{fig:classicalconfigurations}(b), top picture]. Even-membered chains on the other hand are mirror
symmetric about a link and cannot show any local magnetization in the ground state. In this case an alternating quantum
dimerization is more suggestive if $s$ is not very large, where neighboring spins are more strongly correlated on odd
links and less on even links. Another difference is that possible residual spin degrees of freedom at the edges, the
so-called edge spins, first introduced in the context of valence bond ground states and long $s=1$
AFHCs,\cite{AKLT,AKLT2,kennedy1990,Sorensen1994} are effectively coupled ferromagnetically for odd $N$ and
antiferromagnetically for even $N$ (Sec.~\ref{subsec:edgespins}). Notably, rings with periodic boundary conditions
always possess mirror symmetry about a link and a site for both even and odd $N$, so that there is no fundamental
symmetry difference as in chains. For chains the symmetry properties of the lowest multiplets in each sector $S$, which
for $S \ge S_g$ represent the $L$ band, are summarized in Table~\ref{symmetries},\cite{NPK05} in agreement with the
findings for the $s=1/2$ chain in Ref.~[\onlinecite{PhysRevB.46.10866}]. It is also noted that multiplets in a specific
$S$ sector share the spin-flip parity of the corresponding state of the $L$ band, whenever this is a good quantum
number.

\begin{table}
\caption{Symmetry properties of the lowest spin multiplets in each $S$ sector for the AFHC.} \label{symmetries}
\begin{ruledtabular}
\begin{tabular}{cccccc}
$N$ && $s$ && mirror & spin-flip \\
& &&& parity & parity \\
\hline
even && integer or half-integer && $(-1)^{S+Ns}$ & $(-1)^{S+Ns}$ \\
odd && integer $s$, $S \ge S_g$ && $+1$ & $(-1)^{S-s}$  \\
odd && integer $s$, $S < S_g$ && $(-1)^{S-s}$ & $(-1)^{S-s}$  \\
odd && half-integer $s$, $S \ge S_g$ && $+1$ & - \\
odd && half-integer $s$, $S < S_g$ && $(-1)^{S-s}$ & - \\
\end{tabular}
\end{ruledtabular}
\end{table}

%%%%%%%%%%%%%%%%%%%%%%%%%%%%%%%%%%%%%%%%%%%%%%%%%%%%%%%%%%%%%%%%%%
% H_AB Model
%%%%%%%%%%%%%%%%%%%%%%%%%%%%%%%%%%%%%%%%%%%%%%%%%%%%%%%%%%%%%%%%%%
\section{\label{sec:AB} The $H_{AB}$ Model}

For a number of bipartite molecular nanomagnets with antiferromagnetic Heisenberg interactions such as regular even
wheels, modified even wheels, and grid molecules, the lowest lying spectrum was observed to consist of rotational bands
or sets of states whose energies increase as $E(S,q) = \frac{1}{2} \Delta S(S+1) + \epsilon(q)$, with appropriate gap
$\Delta$ and "dispersion relation" $\epsilon(q)$, where $q$ is a suitable index or quantum
number.\cite{And52,Bernu92,HabRing} The lowest band ($L$ band) and the set of higher-lying bands ($E$ band) are
distinguished according to a selection rule and the different nature of the excitations associated to them.  This
structure of the excitations is reminiscent to that shown in Fig.~\ref{fig:spectra} for the AFHCs, and the $E(S)
\propto S(S+1)$ energy dependence is indeed well obeyed for the odd chains, but as was demonstrated in the previous
section pronounced deviations occur in the antiferromagnetic region for the even chains. In order to examine the
even-odd effect the $L \& E$-band picture is therefore reanalyzed.

In the $L \& E$ band picture the $L$ band is described by an effective Hamiltonian $H_{AB}$ of two collective
spins,\cite{HabRing}
\begin{equation}
\label{eq:hab}
  {H}_{AB} = \Delta_{AB} \mathbf{S}_A \cdot \mathbf{S}_B,
\end{equation}
where ${\mathbf{S}}_{A} = \sum_{i \epsilon A}{\mathbf{s}}_i$ and ${\mathbf{S}}_{B} = \sum_{i \epsilon B}{\mathbf{s}}_i$
are the total spin operators of the two sublattices $A = \{1,3,\ldots\}$ and $B = \{2,4,\ldots\}$. The value of
$\Delta_{AB}$ can be determined using symmetry or mean-field
arguments,\cite{bernu,RotationalBands,HabRing,WaldmannMn3x3NVT} and is given for chains and rings in
Eq.~\eqref{eq:DABdef}. The eigenvalues of $H_{AB}$ are easily determined to be
\begin{equation}
\label{eq:ev_hab}
E(S)=\frac{\Delta_{AB}}{2}[S(S+1)-S_A(S_A+1)-S_B(S_B+1)],
\end{equation}
where $|S_A-S_B|\leq S \leq (S_A+S_B)$. For given values of $S_A$ and $S_B$ the energy spectrum consists of rotational
bands. In the $L$ band $S_A$ and $S_B$ assume their maximal values, and the slope is calculated to $g(S) = 1$. The next
higher-lying states of $H_{AB}$, which involve the states where either $S_A$ or $S_B$ is reduced by one, are related to
the $E$ band. However, they do not reproduce the energies of $H$ because the spatial fluctuations, which give rise to
dispersion of these excitations, are not accounted for in the $H_{AB}$ model [that is, $\epsilon(q)$ is not obtained
correctly].\cite{HabRing}

Based on the observation that the energies follow an $S(S+1)$ dependence and that the calculated eigenvalues and
matrix elements become quantitatively exact for large $s$, the $H_{AB}$ model has been called classical or
semi-classical. However, this notion is misleading as $H_{AB}$ retains a quantized spectrum. The basic
underlying assumption of the model is in fact that the correlation length is infinite. The $H_{AB}$ model could
then be regarded as a symmetrized version of the bipartite Heisenberg Hamiltonian, and in fact it is part of the
greater class of the symmetrized effective Hamiltonians which we introduce next.

With a unitary symmetry operator $P$, a translation operator for example, the symmetrized effective Hamiltonians are
constructed as follows: if a Hamiltonian $H$ remains unchanged after $j$ transformations with $P$,
\begin{equation}
P^{-j}H P^{j}=H,
\label{eq:theopj}
\end{equation}
for example if $P^j=1$, then an effective Hamiltonian $H_{\text{eff}}$ is formed by adding up all $j$ transformations
of $H$,
\begin{equation}
H_{\text{eff}}=\frac{1}{j}\sum_{i=0}^{j-1}{P^{-i}H P^{i}}
\label{eq:heffsum}
\end{equation}
This effective Hamiltonian commutes with $P$, since $P^{-1}H_{\text{eff}}P = H_{\text{eff}}$. A common  eigenbasis can
thus be found where the eigenvalues of the unitary $P$ have magnitude 1, and one derives from Eq.~\eqref{eq:heffsum}
\begin{eqnarray}
\label{eq:firstorderzeug}
 \left\langle \Psi_{\text{eff}}^n \mid H_{\text{eff}}\mid \Psi_{\text{eff}}^n \right\rangle
 = \left\langle \Psi_{\text{eff}}^n \mid H \mid \Psi_{\text{eff}}^n \right\rangle,
\end{eqnarray}
since $P^i \left| \Psi_{\text{eff}}^n \right\rangle = \pm 1 \left| \Psi_{\text{eff}}^n \right\rangle$ for any $i$. If
then perturbation theory is applied to a non-degenerate eigenstate $\left| \Psi_{\text{eff}}^n \right\rangle$ using
$H=H_{\text{eff}}+V$, $H_{\text{eff}}$ is found to be exact in first order according to Eq.~\eqref{eq:firstorderzeug}.
This establishes also an upper bound for the ground-state energy of $H$, in any spin sector $S$ if $H$ is SU(2)
invariant, as the higher orders in perturbation theory will result in a negative contribution by the variational
principle.

$H_{AB}$ is obtained if $P$ is a sublattice permutation operator, and it therefore follows that $H_{AB}$ is exact in
first order for non-degenerate energy eigenstates or SU(2) multiplets, respectively, and the $L$ band in particular
[see Figs~\ref{fig:spectra}(a), (b)]. For odd chains and even rings, it suffices to use a translation operator on one
(the bigger) sublattice. $j$ is then the size of the sublattice, and $\Delta_{AB}=2/j$. For even chains the two
sublattices have to interchange.

At this point it may be surprising that odd chains can be described by the $H_{AB}$ model, as the two collective spins
${\mathbf{S}}_{A}$ and ${\mathbf{S}}_{B}$ are necessarily of different size and therefore the problem becomes less
symmetric. However, much more important is the fact that the collective spins obey the mirror symmetry for odd $N$,
which is not the case for even $N$ (where $\mathbf{S}_A$ and $\mathbf{S}_B$ are interchanged by this symmetry
operation). For even chains on the other hand the $H_{AB}$ model would be expected to work well, in contrast to our
findings. In a study on the $N=8$, $s=3/2$ chain it was argued that the $L$- and $E$-band states mix producing
deviations in the energies, which is forbidden in rings due to translational symmetry, at least in first
order.\cite{Bia09-cr8zn} The argument is obviously correct, but does not provide further insight, in particular on the
even-odd effect.

Except for energies with $S \lesssim 2s$ in the even chains, Fig.~\ref{fig:slopesevenodd} could suggest that the
$H_{AB}$ model works otherwise well for chains. However, the slopes which approximate the $L$ band in the ferromagnetic
region in Fig.~\ref{fig:slopesevenodd} are significantly larger than what is predicted by Eq.~\eqref{eq:DABdef}. This
is not easily accounted for within the $H_{AB}$ model. The upper bound given by $H_{AB}$ for the energies in the $L$
band is hence not very tight for both even and odd chains [see Figs.~\ref{fig:spectra}(a) and (b)], giving room to
convex deviations from a $S(S+1)$ dependence, which in the case of the even chains indeed occur for $S \lesssim 2s$.

It is also interesting to inspect the lowest energies in odd chains for $S \leq S_g$ and compare to those of
$H_{AB}$. $H_{AB}$ predicts a linear dependence on $S$, i.e., up to a constant $E(S) = -\frac{1}{2}\Delta_{AB}
(N-1) s S$ or $\delta_E(S) = S_g - S$ for $S \leq S_g$. A linear dependence is indeed observed in this spin
sector as shown in Fig.~\ref{fig:edgeodd}, but the slopes are much lower than what is predicted by $H_{AB}$.
This is attributed to the neglected spatial fluctuations or dispersion of the excitations, which generally
lowers the minimal energy.

For even rings and other bipartite clusters it is well established that the $H_{AB}$ model becomes more valid the
smaller $N$ and/or the larger $s$ is.\cite{HabRing,Chiolero98} Regarding chains, a similar tendency should be expected,
which is {\it not} observed; e.g. the even-odd effect persists for large $s$ as discussed in Sec.~\ref{sec:evenodd}. In
contrast to rings the open boundaries in chains induce inhomogeneities, which are apparently not properly reflected in
a model with infinite correlation length.

Taken together, the $H_{AB}$ model appears appealing at first sight based on its $S(S+1)$ energy dependencies,
which are consistent with the energy spectrum in the AFHCs up to the even-odd effect displayed in
Fig.~\ref{fig:slopesevenodd}, and because it is exact in first-order perturbation theory. However, the detailed
analysis revealed its deficiencies, which appear to be connected to its neglect of spatial fluctuations or the
assumed infinite correlation length.

%%%%%%%%%%%%%%%%%%%%%%%%%%%%%%%%%%%%%%%%%%%%%%%%%%%%%%%%%%%%%%%%%%
% NLSM
%%%%%%%%%%%%%%%%%%%%%%%%%%%%%%%%%%%%%%%%%%%%%%%%%%%%%%%%%%%%%%%%%%

\section{\label{sec:nlsm} Path integral representation: The O(3) nonlinear $\sigma$-model}

The O(3) nonlinear $\sigma$-model (NLSM) has the potential to be suitable for short chains as it is based on the large
spin, semiclassical limit, and can take inhomogeneous states into account. In general, boundaries lead to residual
topological contributions in the effective action for both half-integer and integer spin chains,\cite{Fradkin1988} a
concept that goes beyond the physics the $H_{AB}$ model is capable of describing. In the following we relate the
concepts of the O(3) NLSM with the numerical findings for the spectra of the antiferromagnetic Heisenberg model on
short chains. In the limit of vanishing spatial fluctuations we indeed identify rotational bands in the spectra. For
long chains the topological boundary terms have been interpreted as remaining free spins at the edges of the chain
(edge spins).\cite{NgNLSM,Sorensen1994} We show that this interpretation is not supported by our numerical data and
cannot explain the even-odd effect in the spectra of short AFHCs. On the contrary, we find that in short chains the
alternating magnetization can no longer be separated from the uniform magnetization, as it is assumed in many theories.

\subsection{Derivation of the O(3) NLSM}

The O(3) NLSM originates in an effective path integral representation of the AFHC in the low energy continuum
limit\cite{Haldane, Haldane2} (see  also Appendix \ref{derivation_nlsm}). The Euclidean action of a single isolated
spin $\mathbf{s}_{i}$ with Hamiltonian $H$ is best expressed in terms of spin coherent states reading\cite{Auerbach98}
\begin{eqnarray}\label{action_singlespin}
A [\mathbf{\Omega}]= \int_0^\beta d\tau \langle\mathbf{\Omega}|H|\mathbf{\Omega}\rangle -i s \,\omega[\mathbf{\Omega} ],
\end{eqnarray}
where $\beta = 1/(k_B T)$. States $|\mathbf{\Omega}\rangle$ are labeled by a three-dimensional unit vector
$\mathbf{\Omega}^2=1$ defined by the eigenstate equation $\mathbf{\Omega}\cdot \mathbf{s}_{i}
|\mathbf{\Omega}\rangle=s|\mathbf{\Omega}\rangle$. The kinetic energy of the spin enters in the last term in
Eq.~\eqref{action_singlespin}, defined as
\begin{eqnarray}\label{berryphase}
i s \,\omega[\mathbf{\Omega}]&=&-i s  \int_D d\tau du \, \mathbf{\Omega}
\cdot (\partial_\tau \mathbf{\Omega}\times\partial_u \mathbf{\Omega}),
\end{eqnarray}
where $D$ is the region bounded by the closed curve $\mathbf{\Omega}(\tau)$ on the unit sphere. The auxiliary variables
$u$ and  $\tau$ parametrize $D$. The action is unambiguously defined provided that $2s$ is an integer.\cite{Witten1983}
It should be pointed out that in deriving the action in Eq.~\eqref{action_singlespin} the assumption that
$|\mathbf{\Omega}(\tau+\delta \tau )\rangle-|\mathbf{\Omega}(\tau )\rangle$ is of order $O(\delta \tau)$ is used.
However, there is no physical reason why the difference of two paths at adjacent time steps  should be a small
quantity.\cite{Dai2005} Only for large $s$ the overlap of two spin coherent states becomes infinitely small
\cite{Auerbach98}, and therefore only in this limit the action in Eq.~\eqref{action_singlespin} becomes exact.

For finite chains of length $N$ the action $A$, Eq.~\eqref{action_singlespin}, generalizes to
\begin{eqnarray}
\label{action_manyspin}
 A[\mathbf{\Omega}]=\int_0^\beta d\tau \langle\mathbf{\Omega}|H|\mathbf{\Omega}\rangle-i s\sum_{i=1}^N\omega[\mathbf{\Omega}_i],
\end{eqnarray}
where $\langle\mathbf{\Omega}|H|\mathbf{\Omega}\rangle=J s^2 \sum_{i=1}^N \mathbf{\Omega}_{i} \cdot
\mathbf{\Omega}_{i+1}$ and
$|\mathbf{\Omega}\rangle=|\mathbf{\Omega}_{1},\mathbf{\Omega}_{2},\dots,\mathbf{\Omega}_{N}\rangle$ is a product of
single spin states on each lattice site. In order to derive an effective field theory some assumptions must be made
which are justified in the low energy and large $s$ limit. Since it is expected that at least for short range order the
chain exhibits sizable antiferromagnetic correlations, the local spin field
is decomposed into a N\'eel field $\mathbf{n}$ and a transverse canting field $\mathbf{l}$
\begin{eqnarray}\label{decomposition}
\mathbf{\Omega}_i(\tau)=(-1)^{i+1}\mathbf{n}(x_i,\tau)\sqrt{1-\frac{\mathbf{l}(x_i,\tau)^2}{s^2}}+\frac{1}{s}\mathbf{l}(x_i,\tau),\;\;
\end{eqnarray}
where $x_i=i a$ and $ a$ is the lattice constant. The fields $\mathbf{n}$ and $\mathbf{l}$ are assumed to be slowly
varying and chosen to fulfill $\mathbf{l}(x_i,\tau)\cdot \mathbf{n}(x_i,\tau)=0$ and $|\mathbf{n}(x_i,\tau)|=1$. Field
$\mathbf{l}$ contains the Fourier modes near $0$ and field $\mathbf{n}$ those near $k_0$, the ordering wave vector.
Thus $\mathbf{l}$ roughly represents the net magnetization, which is assumed to be small.

Then Eq.~\eqref{decomposition} is inserted in Eq.~\eqref{action_manyspin}, $L=Na$ is set and the continuum limit is
taken. After some technical calculations (see Appendix~\ref{derivation_nlsm}) the O(3) NLSM is generated:
\begin{eqnarray}\label{nlsm}
A= \int_0^L dx\int_0^\beta d\tau \left[\frac{1}{2 \gamma \upsilon}
\left(\partial_\tau\mathbf{ n}\right)^2+
\frac{\upsilon}{2\gamma}\left(\partial_x\mathbf{n}\right)^2\right]+A_{\rm{top}}\;\;
\end{eqnarray}
with $\upsilon=2Jsa$, $\gamma=2/s$ and
\begin{eqnarray}\label{stop0}
A_{\rm{top}}&=&-i s \sum_{i=1}^N (-1)^{i+1} \omega[\mathbf{n}(x_i)].
\end{eqnarray}
In Eq.~\eqref{nlsm} the field $\mathbf{l}$ has been integrated out giving
$\mathbf{l}=\frac{i}{4J}(\mathbf{n}\times\partial_\tau \mathbf{n})$. In order to understand the topological meaning of
$A_{\rm{top}}$, for even $N$ Eq.~\eqref{stop0} is rewritten as
\begin{eqnarray}\label{stop}
A_{\rm{top}} &=&-i s \sum_{i=1}^{N/2}\omega[\mathbf{n}(x_{2i-1})]-\omega[\mathbf{n}(x_{2i})]\\
&\approx& i \frac{s}{2} \int_0^L dx \frac{\delta \omega}{\delta \mathbf{\Omega}}\mid_{\mathbf{\Omega}
=\mathbf{n}}\cdot \partial_x\mathbf{n}. \label{integral}
\end{eqnarray}
After evaluating the integral in Eq.~\eqref{integral} and taking into account an additional uncompensated Berry phase
for $N$ odd,\cite{Fradkin1988} one generally obtains
\begin{eqnarray}\label{obc_stop}
A_{\rm{top}} &=& i \frac{s}{2} \left\{4 \pi Q +\omega[\mathbf{n}(L)]-(-1)^{N}\omega[\mathbf{n}(0)]\right\},
\end{eqnarray}
where $Q$ is integer valued and simply counts the winding number of the field $\mathbf{n}$.  For periodic boundary
conditions and $N$ even, Eq.~\eqref{obc_stop} reduces to $A_{\rm{top}}=i \theta Q$, the so-called
$\theta$-term,\cite{Haldane,Haldane2} where $\theta=2 \pi s$. $A_{\rm{top}}$ depends only on the topology of the path.

In general, with respect to the derivation of Eq.~\eqref{nlsm} for open boundaries, when performing the continuum limit
for open chains additional boundary terms in the action are expected to be found. In some cases these terms can simply
be included by introducing an effective length. For alternating sums though, the cases of summing over an odd or even
number of terms need to be distinguished. Higher order terms in the bulk action are found which are assumed to be small
[see Eq.~\eqref{boundaryterms2} in Appendix~\ref{derivation_nlsm}].

The topological term $A_{\rm{top}}=i\theta Q$ in Eq.~\eqref{nlsm} is known to play a crucial role for the behavior in
the thermodynamic limit. It enters the path integral with a factor $e^{-i 2\pi s Q}$, thus for integer spin it does not
affect the partition function of the chain, while for half-integer chains $\theta=\pi$ yields an alternating factor
$(-1)^Q$. The renormalization group treatment of the action in Eq.~\eqref{nlsm} with the constraint $|\mathbf{n}|=1$
leads to an increase of $\gamma$ which is at first independent of $A_{\rm{top}}$ at large energy scales. Without the
topological term this leads to a strong coupling fixed point with a mass gap $\Delta_H \sim 0.4Je^{-\pi s}$ for integer
spin chains, which has also been suggested from the calculation of the exact $S$ matrix.\cite{Zamolodchikov1979}
However, with the topological term for $\theta=\pi$ the renormalization behavior is altered and becomes difficult to
analyze at small energy scales, but it is known that half-integer spin chains are massless and belong to the
universality class of the SU(2) Wess-Zumino-Witten model with topological coupling $k=1$ \cite{Affleck1987,Shankar1990,
Zamolodchikov1992, Bietenholz1995} in agreement with the Lieb-Schultz-Mattis theorem \cite{lieb1961} and the Bethe
ansatz solution for the integrable spin-$\frac{1}{2}$ chain. This difference between integer and half-integer spin
chains has become well-known in the literature.\cite{Haldane,Haldane2} However, in the small chains that are considered
in this work, the renormalization group flow is terminated by the finite length of the system at rather large energy
scales, which exceed the mass gap, i.e., before $A_{\rm{top}}$ makes a significant difference. Therefore, there is no
need to distinguish between massive and massless theories in the following analysis.

\subsection{The Rotational Band in the O(3) NLSM}

In order to connect with the $H_{AB}$ model, spatial fluctuations in the action are initially ignored. The problem then
reduces to a $(0+1)$-dimensional one and the action of Eq.~\eqref{nlsm} taking into account Eq.~\eqref{obc_stop} is
given by
\begin{eqnarray}\label{action_rigidrotator}
A= \frac{I}{2}\int_0^\beta d\tau \,(\partial_\tau\mathbf{n})^2+i \epsilon \,s \,\omega[\mathbf{n}],
\end{eqnarray}
where $I=N/4J$, $\epsilon=0$ for $N$ even and $\epsilon=1$ for $N$ odd. For $\epsilon=0$ this yields the path integral
of the $3$-dimensional rigid rotator\cite{grosche1987}, with  energy levels
\begin{eqnarray}\label{rotationalbandeven}
E(S)=\frac{1}{2 I} S(S+1) \quad\mbox{($N$ even)},
\end{eqnarray}
where $S=0,1,\dots$ and each energy level is $(2S+1)$-fold degenerate.

The case $\epsilon=1$ can also be solved exactly, e.g., in the CP$^1$ representation by introducing an independent
gauge field.\cite{Rabinovici1984} This auxiliary field can be set to zero, and the Lagrangian becomes the rigid rotator
Lagrangian again. However, the variation of the gauge field generates a constraint on the possible quantum numbers. One
obtains
\begin{eqnarray}\label{rotationalbandodd}
E(S)=\frac{1}{2 I} \left[ S(S+1)-s^2 \right]\quad\mbox{($N$ odd)}
\end{eqnarray}
with the allowed $S$ now constrained to $S=s,s+1,\dots$.

For open boundaries there is an additional discrete lattice symmetry $i\rightarrow N+1-i$, which is maintained by
introducing an effective length $L= (N+1)a$ in the continuum model. The moment of inertia then reads $I=(N+1)/4J$, and
thus Eqs.~\eqref{rotationalbandeven} and \eqref{rotationalbandodd} recover the slopes of the $H_{AB}$ model in
Eq.~\eqref{eq:DABdef} up to order $O(1/N^2)$. Taking into account short range fluctuations in the Berry phase of a
single spin \cite{Dai2005} or higher order operators in the effective action  generally renormalizes the coupling
constants. A similar mapping of magnetic molecular rings onto a rigid rotator model has been performed in
Refs.~[\onlinecite{Normand2001,Maier2001}].

Neglecting spatial fluctuations in the O(3) NLSM apparently renders its physics equivalent to the $H_{AB}$ model.
However, if spatial fluctuations are included, the coupling to the rotational band has to be analyzed, which will alter
Eqs.~\eqref{rotationalbandeven} and \eqref{rotationalbandodd}. In particular, one would expect the spatial derivative
term in Eq.~\eqref{nlsm} to become more important once the total spin is excited, since the additional net spin needs
to be distributed along the chain. A discussion of this point with regards to the spin density will be given in
Sec.~\ref{sec:spindensity}. However, as soon as the spatial derivative term is included in Eq.~\eqref{nlsm} and the
full Lagrangian is treated the spectra cannot be extracted as easily anymore, but it is possible to consider the
simplified case of edge excitations.

\subsection{Effective Edge-Spin Hamiltonian}\label{subsec:edgespins}

In Ref.~[\onlinecite{Sorensen1994}] long $s=1$ chains of even or odd length were modeled by the O(3) NLSM coupled to
$s_{edge}=\frac{1}{2}$ edge spins. The constraint $|\mathbf{n}|^2=1$ of the nonlinear $\sigma$-field was relaxed by
adding an artificial mass term and a repulsive $\lambda \mathbf{n}^4$ interaction.  On a mean field level the parameter
$\lambda$ can be assumed to be small. Then the field $\mathbf{n}$ can be integrated out, and an effective Hamiltonian
where only the edge spins couple to each other results\cite{Sorensen1994}
\begin{eqnarray}\label{Hedge}
H_{edge} =  J_{\rm eff} \mathbf{s}_1^\prime \cdot \mathbf{s}_L^\prime,
\end{eqnarray}
where $\mathbf{s}_1'$ and $\mathbf{s}_L'$ are spin operators representing the edge spins. Equation~\eqref{Hedge} is a
valid approximation at energies much smaller than the Haldane gap, for chains much longer than the correlation length.
The effective exchange interaction $J_{\text{eff}} \sim (-1)^N e^{-N/\xi} J$ between the edge spins is ferromagnetic
for $N$ odd and antiferromagnetic for $N$ even, where $\xi$ is the spin-spin correlation length of the corresponding
spin chain. The edge-spin picture hence gives rise to a pronounced even-odd difference in the $L$-band energy spectrum
at small values of $S$. However, this difference can only be derived for long chains, in contrast to the even-odd
effect observed in Sec.~\ref{sec:evenodd}, which is a property of chains that are shorter than the correlation length.

For long chains the existence of edge states is not restricted to integer spin. Based on Eq.~\eqref{obc_stop} and
interpreting the residual Berry phase with free spins, edge spins of magnitude $s_{edge}=\frac{s}{2}$ for $s$ integer
and $s_{edge}=\frac{1}{2}(s-\frac{1}{2})$ for $s$ half-integer have been proposed.\cite{NgNLSM} Half-integer spin
chains have thereby been pictured as a continuum model of a spin-$\frac{1}{2}$ chain coupled to two "impurity" spins of
magnitude $\frac{1}{2} (s-\frac{1}{2})$. Equation~\eqref{Hedge} then predicts the following spectrum: for $N$ even the
ground state has $S_g=0$ and the edge spins form a singlet. The lowest energy excitations can  be constructed by
exciting the two edge spins into an $S=1,2,\dots,2s_{edge}$ state, and the excitation energies are given by
\begin{eqnarray}\label{spectrum_edge_even}
E(S)-E(0)= J_{\rm eff} S(S+1) \quad (\text{$N$ even}).
\end{eqnarray}
For $N$ odd, due to the ferromagnetic effective coupling $J_{\rm eff}<0$ the lowest energy is obtained if the edge
spins are coupled to their maximal value $2 s_{edge}$ (in case of half-integer spin the bulk spin-$\frac{1}{2}$
chain contributes the additional spin $\frac{1}{2}$ to the total spin $S_g=s$ of the ground state). Excitations
with $S<S_g$ are constructed by coupling the edge spins to lower spin values $2 s_{edge}-S^\prime$ where $S' =
S_g - S$. The corresponding excitation energies are
\begin{eqnarray}
\label{spectrum_edge_odd}
E(S_g - S')-E(S_g)=|J_{\rm eff}| S'(S'+1) \quad (\text{$N$ odd})
\end{eqnarray}
for $S'=1,2,\dots,2s_{edge}$. For large system sizes the edge-spin picture has been numerically verified.
\cite{kennedy1990,Sorensen1994,LouQinNgSu,LouQinChen,QinNgSu} However, for short system sizes  it does not seem
justified to regard the edge states decoupled from the bulk states, as explained above.

The coupling of spatial fluctuations to the bulk states for shorter system sizes may be described by a more general
effective edge-spin Hamiltonian with the following ansatz $H' = \frac{\mathbf{l}^2}{2I}+H_I$ where $H_I=\lambda_u
\mathbf{l} \cdot (\mathbf{s}_1^\prime + \mathbf{s}_L^\prime) + \lambda_s\mathbf{n} \cdot [\mathbf{s}_1^\prime - (-1)^N
\mathbf{s}_L^\prime]$, which is appropriate if the coupling to the edge spins is weak. In case the two parameters
$\lambda_u$ and $\lambda_s$ are both sufficiently small the field $\mathbf{l}$ can be integrated out, which recovers
the edge-state Hamiltonian in Eq.~\eqref{Hedge}. However, it is so far unclear how to treat the effective model $H'$
for the general case of short chains in order to extract the modified spectrum, which remains a task for future
research.

\begin{figure}
\includegraphics[width=7cm]{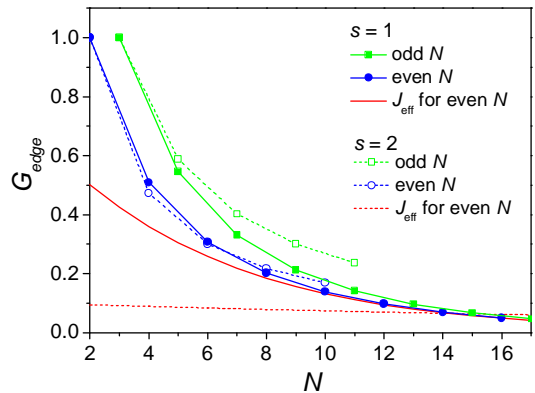}
\caption{\label{fig:gAFgEvsN}(Color online)
Slopes $G_{edge}$ for odd (green squares)  and even (blue circles) $N$. Solid symbols are for chains
with $s=1$, and open ones with dashed lines  for chains with $s=2$.  The red  lines  represent the slopes
$G_{edge}=J_{\text{eff}}$ for large even $N$ resulting from Eq.~(\ref{spectrum_edge_even}) with
$J_{\text{eff}}=0.35 e^{-N/6}J$ for $s=1$  and $J_{\text{eff}}=0.05 e^{-N/33}J$ for $s=2$ according to
Ref.~[\onlinecite{QinNgSu}]. The solid line is for $s=1$, and the dashed one for $s=2$.
Lines are guides to the eye. }
\end{figure}

In Refs.~[\onlinecite{LouQinNgSu,QinNgSu}] it has been conjectured on the basis of numerical density matrix
renormalization group (DMRG) data that the edge-state picture stays robust up to very small $N$, even when $N$ becomes
smaller than the correlation length. In our data, however, we do not see signatures of edge states. First, for odd
chains the lowest excitations $E(S)$ for $S<S_g$ scale rather linearly with $S$ in contrast to the quadratic behavior
predicted in Eq.~\eqref{spectrum_edge_odd} (see Fig.~\ref{fig:edgeodd} and the corresponding discussion in
Sec.~\ref{sec:evenodd}). Interestingly, a linear dependence results from the $H_{AB}$ model. Second, for even chains
the excitations at small $S$  generally do not obey $E(S)\propto S(S+1)$ which contradicts the edge-spin prediction of
Eq.~\eqref{spectrum_edge_even}. In particular, the deviation from quadratic behavior is observed up to $\sim$2$s$,
while edge states would correspond to the lowest $s$ excitations [see Fig.~\ref{fig:slopesevenodd}(a)]. The edge states
cannot be distinguished from the higher excitations up to $\sim$2$s$. Finally, the scaling with chain length $N$ of the
first excited state above the ground state does not follow the exponential decay of the edge-spin picture. To
illustrate this let us define the corresponding (unnormalized) slopes in the $E$ vs $S(S+1)$ diagram:
\begin{equation}
G_{edge} = \left\{
  \begin{array}{l l}
    -\dfrac{\Delta_{AB}}{2} g(s)& \quad \text{($N$ odd)},\\ % & \\
    \dfrac{\Delta_{AB}}{2} g(1) & \quad \text{($N$ even)},
  \end{array} \right.
\end{equation}
where $g(S)$ is defined in Eq.~(\ref{def:normalizedslope}). For $N$ even,  $G_{edge}$ is related to $g_{AF}$  by
$G_{edge}=g_{AF}\Delta_{AB}/2$ (see also Fig.~\ref{fig:gAF}), and for $N$ odd, it is related to $\delta_E$ by
$G_{edge} = (N-1) \Delta_{AB}\delta_E(s-1)/4$ (see also Fig.~\ref{fig:edgeodd}). Fig.~\ref{fig:gAFgEvsN} shows
$G_{edge}$ for $s=1$ and 2 for even and odd  $N$. According to Eqs.~(\ref{spectrum_edge_even}) and
(\ref{spectrum_edge_odd}), for long chains these slopes are given by $G_{edge}=J_{\text{eff}}$ for $N$ even and
$G_{edge}=|J_{\text{eff}}|/s$ for $N$ odd. The fit curves to the data for long even chains which thus have
the form $G_{edge}=0.35 e^{-N/6}J$ for $s=1$ and $G_{edge}=0.05 e^{-N/33}J$ for $s=2$ where the fit parameters
are taken from Ref.~[\onlinecite{QinNgSu}] are also plotted. The deviations for smaller $N$ are obvious.
Furthermore, the deviations appear to increase with increasing $s$, demonstrating that the edge-spin picture
becomes less appropriate with increasing $s$, in contrast to the observations for the even-odd effect of
Sec.~\ref{sec:evenodd}, which is present even for large $s$.

In conclusion, the standard edge-spin picture cannot account for the even-odd effect. Including couplings of edge spins
to both $\mathbf l$ and $\mathbf n$ in the NLSM would result in a more complete model that may remedy the situation.
However, this would imply that uniform and alternating magnetization become strongly coupled and can no longer be
treated separately.

%%%%%%%%%%%%%%%%%%%%%%%%%%%%%%%%%%%%%%%%%%%%%%%%%%%%%%%%%%%%%%%%%%
% Spin 1/2
%%%%%%%%%%%%%%%%%%%%%%%%%%%%%%%%%%%%%%%%%%%%%%%%%%%%%%%%%%%%%%%%%%
\section{\label{sec:spinhalf} Comparison to the spin 1/2 chain}

As both the $H_{AB}$ model and the NLSM fail to account for the even-odd effect quantitatively, it may be instructive
to turn to the special case of $s=1/2$ chains to examine if quantum effects play an important role. The spectrum of
finite $s=1/2$ chains is quantitatively very well understood, not only from the Bethe ansatz,\cite{bortz} but also in
terms of effective bosonic quantum numbers from bosonization,\cite{AffleckGepnerSchulzZiman89,PhysRevB.46.10866} which
establishes the $s=1/2$ chain as an excellent reference.

The description of the spectrum of the $s=1/2$ chain in terms of bosonic quantum numbers
\cite{bortz,AffleckGepnerSchulzZiman89,PhysRevB.46.10866} results in an almost equally spaced energy spectrum in the
form of a conformal tower. There are corrections of order $1/N^2$ and $1/(N\ln N)$ to the spectrum, but this effective
description works well for $N \agt 10$. An $L$ band can also be observed, except that in this case the lowest lying
energy states of a given $S$ are created by adding bosonic particles with zero momentum, and the number of bosonic
particles is given by the $S^z$ quantum number, the projection of $S$ along the $z$ axis. For $s=1/2$ chains with $N$
both even and odd the excitation energies in the $L$ band are given by\cite{PhysRevB.46.10866}
\begin{equation} \label{even-spinhalf}
E(S) =  \frac{\pi v}{N+1} S^2
\end{equation}
up to higher order corrections in $1/N^2$ and $1/(N\ln N)$, where $v=\pi J/2$. Hence, there is no even-odd effect to
lowest order in the excitation spectrum. There is a contribution to the ground state energy of order $1/N$, which is
positive for odd $N$ and negative for even $N$,\cite{anfuso2003} but this is not related to the even-odd effect in
Sec.~\ref{sec:evenodd}. It is important to notice that in the $s=1/2$ chain the $E(S)$ dependence is predicted to be
changed from the $S(S+1)$ behavior to a simple $S^2$ behavior, analogously to the charging energy of a capacitor, which
shows a quadratic energy dependence in the charge $Q^2$.

\begin{figure}
\includegraphics[width=7cm]{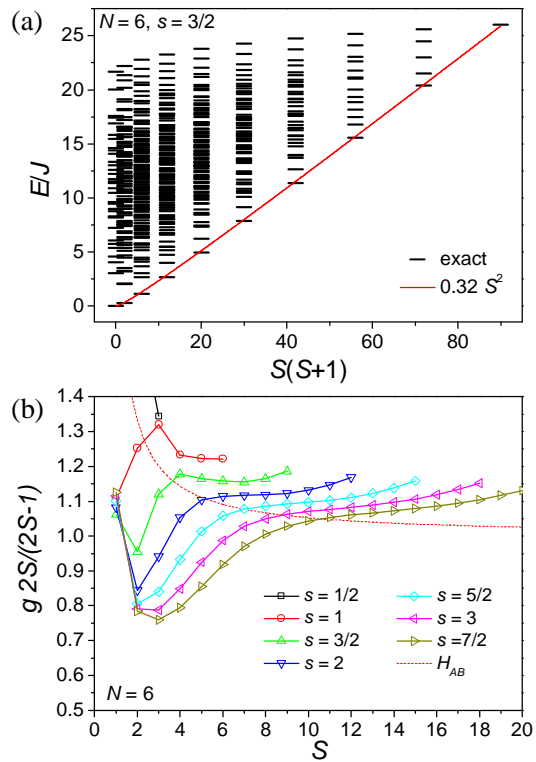}
\caption{\label{fig:s-squared} (Color online)
(a) Energy spectrum of the $N=6$, $S=3/2$ chain as function of $S(S+1)$. The (red) line is the fit $E(S) = 0.32 S^2$.
(b) Slopes $g(S)\times 2S/(2S-1)$ as functions of $S$ for the $N=6$ chain for different spins $s$ (symbols). The dashed
(red) line is the prediction of the $H_{AB}$ model. }
\end{figure}

Interestingly, such an $S^2$ behavior seems to agree better with the $L$-band energies for even chains with larger $s$,
as shown in Fig.~\ref{fig:s-squared}(a) for the case of $N=6$ and $s=3/2$ corresponding to the Cr$_6$ molecule. The
$S^2$ behavior is consistent with the entire $L$ band with $\frac{\pi v}{N+1}=0.32J$ or $v\approx 0.71J$ according to
the fit in the figure. In order to test this further, the slope of the $L$ band in an energy vs. $S^2$ diagram, which
is given by $g(S) \times 2S/(2S-1)$, is plotted in Fig.~\ref{fig:s-squared}(b) for $N=6$ and different $s$. The $S^2$
dependence doesn't account for the even-odd effect, which in Fig.~\ref{fig:s-squared}(b) is observed as the pronounced
"dip" at small $S$, but provides a significantly better average fit through the spectra compared to the $S(S+1)$
behavior in Fig.~\ref{fig:slopesevenodd}(a). However, this "success" of an $S^2$ behavior does not explain why such an
approach fails for odd chains, which obviously remain to be well described by the $S(S+1)$ behavior as shown in
Fig.~\ref{fig:slopesevenodd}(b), nor does it give an independent estimate for $\frac{\pi v}{N+1}$.

For the $s=1/2$ chain it is also possible to predict local expectation values, such as the alternating
magnetization\cite{PhysRevLett.75.934,PhysRevB.62.4370,PhysRevLett.89.047202} and the dimerization along the chain. For
example the alternating spin expectation values in the $z$ direction can be calculated for the highest weight states in
the $L$ band\cite{PhysRevLett.89.047202}
\begin{eqnarray}
\langle s^z_i \rangle &
 \propto & (-1)^{i+1}\frac{\sin (2\pi S^z x_i/L)}{\sqrt{L \sin (\pi x_i/L)}},
\label{localsz}
\end{eqnarray}
where $L=(N+1)a$ is the effective length. Possible multiplicative corrections\cite{AffleckGepnerSchulzZiman89} of order
$1/\ln L$ and higher order terms have been neglected here. The alternating order always decreases $\propto\sqrt{i}$
near the edges.\cite{PhysRevLett.75.934} The calculation also implies an even-odd effect in the density: For the ground
state of odd $N$ chains with $S^z=1/2$ there is a maximum $\propto 1/\sqrt{L}$ in the middle of the chain, while the
alternating order is zero for even $N$ chains with $S^z=0$. The result shows explicitly that inhomogeneities are
present and important over the entire chain, which were of course neglected in the $H_{AB}$ model. Spin densities for
higher $s$ cases will be discussed in Sec.~\ref{sec:spindensity}.

A similar calculation yields the alternating part of the nearest-neighbor correlation for states in the $L$ band, which
is dimerized
\begin{eqnarray}
\langle \mathbf{s}_i \cdot \mathbf{s}_{i+1}\rangle &
 \propto & (-1)^{i}\frac{\cos (2\pi S x_i/L)}{\sqrt{ L \sin (\pi x_i/L)}}.
\label{dimerization}
\end{eqnarray}
The dimerization becomes very strong and length independent at the edges $\langle \mathbf{s}_1 \cdot
\mathbf{s}_{2}\rangle \sim -0.65$ , but remarkably there is no pronounced even-odd difference, since the cosine
function near the boundary is independent of $S$ being integer or half-integer. Corrections to the edge dimerization
are small down to very short $s=1/2$ chains and the correlation of the first two sites $\langle \mathbf{s}_1 \cdot
\mathbf{s}_{2}\rangle$ is much enhanced compared to the bulk value of $-0.4431$ both for even and odd $N$, despite the
fact that only even chains could potentially lock into a dimerized ground state, while odd chains naively should not be
able to support such a valence bond state. In fact, the difference of the expectation value $\langle \mathbf{s}_1 \cdot
\mathbf{s}_2\rangle$ is only about 15\% between chains of $N=4$ and $N=5$.

In conclusion, the analysis in this section shows that the even-odd effect described in Sec.~\ref{sec:evenodd} is not
present in the quantum theory of the $s=1/2$ chain. An even-odd difference in the spin density can be
observed,\cite{PhysRevLett.89.047202} but this is related to symmetry properties, as also discussed later in
Sec.~\ref{sec:spindensity}. However, strong inhomogeneities are observed in the chain and quantum effects cause the $L$
band to be better described on average by a "charging energy" of the form $E(S)\propto S^2$.

%%%%%%%%%%%%%%%%%%%%%%%%%%%%%%%%%%%%%%%%%%%%%%%%%%%%%%%%%%%%%%%%%%
% Classical
%%%%%%%%%%%%%%%%%%%%%%%%%%%%%%%%%%%%%%%%%%%%%%%%%%%%%%%%%%%%%%%%%%
\section{\label{sec:classicallimit} Classical Antiferromagnetic Heisenberg Chain}

The classical AFHC, where the spins in Eq.~\eqref{eq:Hchain} are treated as classical objects (vectors), is a good
approximation of the quantum model for large $s$. It ignores quantum fluctuations but fully retains spatial
fluctuations, and thus allows to study the importance of the latter. This model has already been studied within the
context of artificial nanostructures,\cite{Lounis08,Politi09} and it has been found that even chains always have a
coplanar and non-collinear ground state. For odd chains the situation is the same, except that for magnetic fields
below a critical field the lowest energy configuration is ferrimagnetic, and an analytical expression for the critical
field was found.\cite{Politi09} This difference reflects the different total ground-state spins $S_g=0$ and $S_g=s$ for
even and odd chains. In this section we present the lowest energy, spin density and nearest-neighbor correlation
functions of the classical AFHC, and compare them to the quantum results, as functions of the normalized squared total
spin
\begin{equation}
S_n^2 = \frac{S(S+1)}{S_f(S_f+1)}
\end{equation}
or $S_n = S/S_f$ in the classical case, where $S_f=Ns$.

The classical analog of Eq.~\eqref{eq:Hchain} is constructed by introducing unit vectors $\mathbf{e}_{i}=
\mathbf{s}_i/s$, whose components commute in the limit $s \to \infty$.\cite{Fis64-class,Joy67-class} The classical
vectors can then be parameterized in spherical coordinates as $\mathbf{e}_i=(\cos\phi_i \sin\theta_i, \sin\phi_i
\sin\theta_i, \cos\theta_i)$. Substitution in Eq.~\eqref{eq:Hchain} minimizes the energy when $\theta_i=
\theta_{N+1-i}$ and the nearest-neighbor relative azimuthal angles $\phi_{i+1}-\phi_i=\pi$,\cite{Coffey92,NPK07} thus
the spin configurations are planar as expected. The classical Hamiltonian then reads
\begin{equation}
H = J s^2 \sum_{i=1}^{N-1}\cos(\theta_i+\theta_{i+1}),
\label{eq:classicalHamiltonian}
\end{equation}
where $\theta_i\in [0,\pi]$ for all $i$. Minimization of Eq.~\eqref{eq:classicalHamiltonian} gives the absolute ground
state.\cite{Coffey92,NPK07,NPK12} Employing rotational symmetry, the lowest energy for arbitrary total magnetization
$\mathbf{S} = s \sum_i \mathbf{e}_i$ can be calculated by adding an external magnetic field term $H_B = - B s \sum_i
\cos\theta_i$ in Eq.~\eqref{eq:classicalHamiltonian}, where $B$ is directed along the $z$ axis (the field is measured
in units of $2\mu_B$ in this section). The direction of the magnetization coincides then with the direction of the
field and $S = s \sum_i \cos\theta_i$. By tuning $B$ the zero-field energies $E(S)$ can be calculated for all values of
$S$ by subtracting the magnetic energy at the end. For odd chains, configurations with magnetization less than the
value of the absolute ground state $S<S_g=s$ are not accessible this way, as $E(S)$ decreases as function of $S$ in
this regime. The calculation of these states is performed by adding a term $H_K=K S^2$ with $K>0$, which favors states
with minimal $S$.

\begin{figure}
\includegraphics[width=7cm]{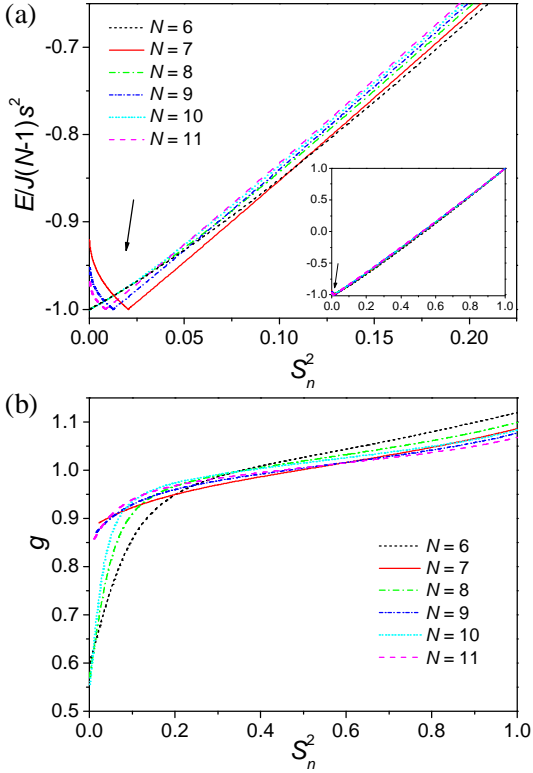}
\caption{\label{fig:classicalslopes} (Color online)
(a) $L$-band energies $E(S)$ of the classical AFHC scaled with $J(N-1)s^2$ for chains with lengths $N=6$ to 11 for
smaller values of $S_n^2$. For the odd chains the energies with $S < s$ do not belong to the $L$ band but to
configurations with magnetization less than the one of the absolute ground state, and are included here for
completeness. The arrow points towards the "kinks" in the energies at the fields where $S = s$. The inset shows the
same figure for the whole $S_n^2$ range. (b) The slopes $g$ of the $L$ band of the classical AFHC as function of
$S_n^2$ for chain lengths $N$ ranging from 6 to 11.}
\end{figure}

\subsection{Slopes and Energies}

The slope in the classical case is determined as $g(S)$ = $\frac{2}{\Delta_{AB}} \partial E/\partial (S^2) =
\frac{1}{\Delta_{AB} S} \partial E/\partial S$, and is related to the inverse magnetic susceptibility $(S/B)^{-1}$:
According to the Legendre transformation $E(S) = E_B +  B S$, the magnetization $S$ is given as $S(B) = -
\partial E_B/\partial B$ (at $T = 0$) and the field as $B(S) = \partial E(S)/\partial S$.
One thus finds
\begin{equation}
\label{hS}
B(S) = \Delta_{AB} S g(S).
\end{equation}
The inverse of this equation gives the magnetization as a function of field, $S(B)$, which implies that $\Delta_{AB}
g(S)$ is the reciprocal susceptibility $(S/B)^{-1}$ (see also Sec.~\ref{sec:evenodd}).

\begin{figure}
\includegraphics[width=7cm]{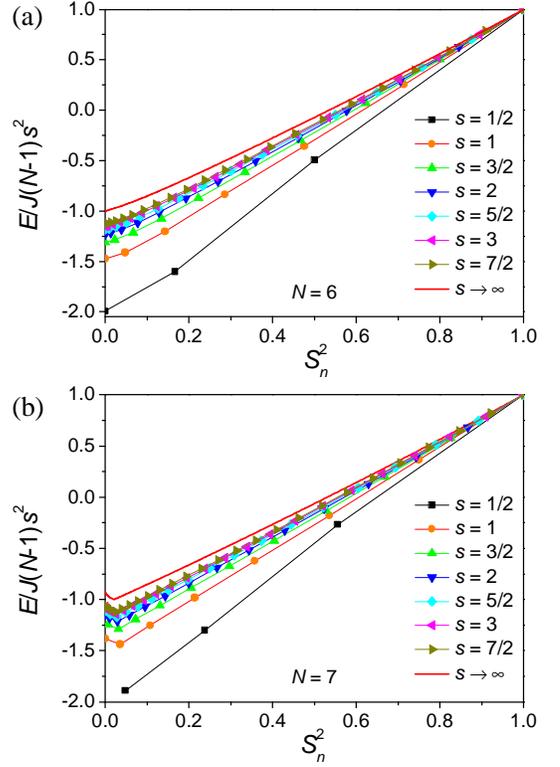}
\caption{\label{fig:n6n7spectra} (Color online)
$L$-band energies $E(S)$ scaled with $J(N-1)s^2 $ for the (a) $N=6$ and (b) $N=7$ chain, for $s$ ranging from
1/2 to $\infty$ (symbols). Results are shown as function of $S_n^2$. For $N=7$ (odd chains) the energies with $S < s$ do not
belong to the $L$ band but to configurations with magnetization less than the absolute ground state, and are
included here for completeness.}
\end{figure}

\begin{figure*}
\includegraphics[width=17.5cm]{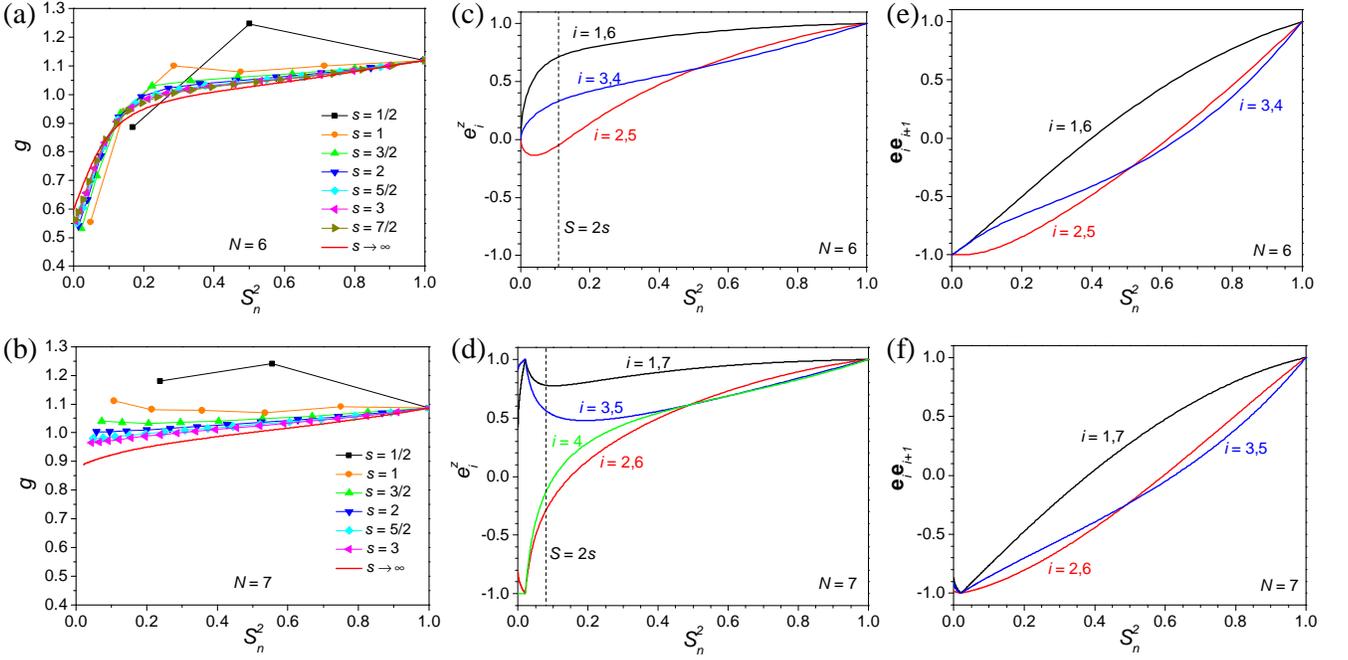}
\caption{\label{fig:n6n7classical} (Color online) Behavior of the classical AFHC for $N=6$ (top) and $N=7$ (bottom) as
function of $S_n^2$. (a,b) The classical slopes where $s \to \infty$ (red lines), along with the slopes of the quantum
AFHC (symbols) of Fig.~\ref{fig:slopesevenodd} are shown. (c,d) Spin density $e^z_i=\cos\theta_i$ along the direction
of total spin $S$. The numbering of the spins start at the edge of the chain. The spin density is mirror-symmetric with
respect to the center of the chain. (e,f) Nearest-neighbor correlation functions $\mathbf{e}_i \cdot \mathbf{e}_{i+1}$.
They are mirror symmetric with respect to the center of the chain.}
\end{figure*}

The $L$-band energies of chains with lengths $N=6$ to $11$ are displayed in Fig.~\ref{fig:classicalslopes}(a) as
functions of $S_n^2$, and the corresponding slopes are presented in Fig.~\ref{fig:classicalslopes}(b). For even chains,
the classical slope $g$ is small at small $S$ and increases rapidly with increasing $S$ (which comes about effectively
like an increase of the external magnetic field $B$ and can be thought of in these terms). In contrast, the slope (for
$S \ge S_g$) for the odd chains starts off at a higher value compared to the even chains, and shows a significantly
weaker dependence on the total magnetization. The slopes for both the even and odd chains become comparable and weakly
varying for $S_n^2 \approx 4/N^2$ or $S \approx 2s$. In Fig.~\ref{fig:n6n7spectra} the energy spectra $E(S)$ of the
quantum AFHC, scaled with the energy of the ferromagnetic state, $J(N-1)s^2 $, are shown for the $N=6$ and $7$ chains
for $s$ ranging from 1/2 to 7/2, and the classical results are also shown. The corresponding slopes $g(S)$ are
presented in Figs.~\ref{fig:n6n7classical}(a), (b). The quantum energies and corresponding slopes approach the
classical ones with increasing $s$. Although convergence is relatively slow, in both the even and odd chains the
classical and quantum slopes exhibit very similar features. Most importantly, the strong down-bending in the slope for
the even chain at small values of $S_n^2$ (or $S$), which is the hallmark of the even-odd effect of
Sec.~\ref{sec:evenodd}, is also present in the classical system. This implies that spatial inhomogeneities must be the
leading mechanism of the even-odd effect, while quantum fluctuations give quantitative corrections to this phenomenon.

\subsection{Local Magnetization and Correlations}

To study the spatial inhomogeneities further local quantities are considered. The local magnetizations of the spins
along the direction of $\mathbf{S}$ (with $S>0$) or spin densities $e^z_i = \cos\theta_i$ and the nearest-neighbor
correlation functions $\mathbf{e}_i \cdot \mathbf{e}_{i+1}$ are shown in Figs.~\ref{fig:n6n7classical}(c), (e) and
\ref{fig:n6n7classical}(d), (f) for $N=6$ and $7$ respectively. The related spin configurations are presented in
Fig.~\ref{fig:classicalconfigurations} for different values of $S_n^2$. $e^z_i$ and $\mathbf{e}_i \cdot
\mathbf{e}_{i+1}$ obey the mirror symmetry of the chain. In even chains the local magnetization $e^z_i$ is zero
everywhere in the ground state as the spins align perpendicular to $\mathbf{S}$. For very small $S$ (or magnetic field
$B$) the $e^z_i$ are almost perpendicular to $\mathbf{S}$, to optimally preserve their exchange energy. With increasing
$S$ the outer spins $e^z_1$ and $e^z_N$ have the largest projection on $\mathbf{S}$ among all spins and gain the most
magnetic energy, while their nearest neighbors $e^z_2$ and $e^z_{N-1}$ turn against the field. This configuration
allows a net magnetization at low exchange energy cost, and the edge spins are in fact very quickly magnetized, which
implies a large magnetic susceptibility or a small slope $g$. The situation is quite different for odd chains since the
ground state is in a ferrimagnetic configuration and the outer spins are already fully aligned with the total
magnetization $S=s$. In order to magnetize the chain further the spins on the odd sites decrease their local
magnetization $e^z_i$, and this allows the spins on the even sites to increase their magnetization. This magnetization
process costs more energy and is less efficient than in the even chain. Hence the susceptibility is smaller and the
slope $g(S)$ is larger for odd chains. This is also reflected in the local correlations in
Figs.~\ref{fig:n6n7classical}(e), (f) where the correlation $\mathbf{e}_1 \cdot \mathbf{e}_{2}$ of the first bond
increases more strongly for odd chains than for even chains, which in turn requires more energy. The markedly different
slopes at small $S$ in the even and odd chains are hence related to the high susceptibility of the outer spins in the
even chains towards magnetic fields, providing an intuitive picture of the even-odd effect.

Remarkably, at $S_n^2 \approx 4/N^2$ or $S\approx 2s$ the differences in correlations and local magnetizations between
even and odd chains start to disappear, and the slopes $g(S)$ in Fig.~\ref{fig:n6n7classical}(a) and (b) become
comparable [see also Fig.~\ref{fig:classicalslopes}(b)]. This crossover region $S \approx 2s$ is marked by a vertical
dotted line in Figs.~\ref{fig:n6n7classical}(c),(d) and the corresponding spin configurations are depicted in
Fig.~\ref{fig:classicalconfigurations}. For $S$ larger than $2s$ the interior spins exhibit nearly identical $e^z_i$
for both even and odd chains, while the outer spins have larger local magnetization.

\subsection{Analytical Results}

Further insight is provided by the analytic calculation of the local magnetization along the chain in the limits $NB
\to 0$ and $NB \to \infty$, respectively. Using a small angle expansion of Eq.~(\ref{eq:classicalHamiltonian}) for
infinitesimal fields $B$ we arrive at a set of coupled equations, which can be solved analytically for the local
magnetization of classical chains with arbitrary even $N$
\begin{equation}
\label{local-mag}
  e^z_i= \frac{B}{4Js}\left[ 1 + (-1)^i \left(2 i - 1-N\right) \right]
\end{equation}
for $i \le N/2$ ($e^z_{N+1-i} = e^z_{i}$ due to mirror symmetry). The local magnetization hence decays linearly from
the edges into the chain. The total magnetization is obtained as $S= NB/(2J)$, where it should be noted that both the
uniform and alternating parts in Eq.~\eqref{local-mag} contribute equally to it. This is remarkable, since it implies
that the alternating part due to the boundary condition is affecting a thermodynamic quantity, i.e., the edge effect is
of order $N$ and cannot be extracted from standard finite-size scaling. The energy is $E(S)=-(N-1)J+(J/N)S^2$. For the
slope thus holds
\begin{equation}
\label{smallg}
g(S\to 0)= \frac{2 J}{N \Delta_{AB}}  = \frac{1}{2} \frac{N}{N-1}.
\end{equation}
This is almost a factor 2 smaller than the prediction $g = 1$ of the $H_{AB}$ model or the slope $g_F$ in the
ferromagnetic region $S_n \to 1$ [Eq.~\eqref{eq:gFexact}], and thus explains the strong reduction of $g(S)$ for even
chains analytically (but does not explain a crossover at $S \approx 2s$). For odd chains we could not find a closed
analytical solution of the linearized equations.

\begin{figure}
\includegraphics[width=8.4cm]{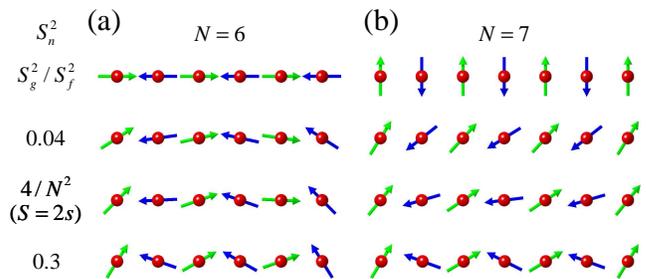}
\caption{\label{fig:classicalconfigurations} (Color online)
Spin configurations of the classical AFHC for chains with lengths (a) $N=6$ and (b) $N=7$ for different values of
$S_n^2$ extracted from Fig.~\ref{fig:n6n7classical} (c) and (d).}
\end{figure}

\begin{figure}
\includegraphics[width=7cm]{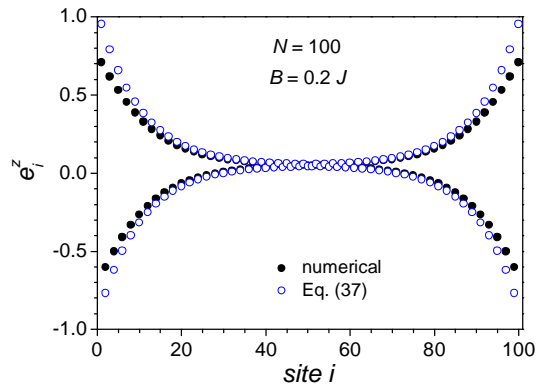}
\caption{\label{fig:local-mag} (Color online)
The local magnetization $e^z_i$ along an even classical chain of length $N = 100$ for a field of $B = 0.2 J$ (solid
symbols) compared to the theoretical prediction in Eq.~\eqref{finitefield} (open symbols).}
\end{figure}

The local magnetization can also be calculated approximately from an effective hydrodynamic theory in a semi-infinite
chain in finite fields (i.e. $NB \gg Js$),\cite{Eggert2007} which can also be derived from the classical version of the
NLSM in Sec.~\ref{sec:nlsm}. This results in the following expression for the local magnetization\cite{Eggert2007}
\begin{eqnarray}
e^z_i  & = & \frac{B}{4Js} - (-1)^i 4 J B s \int_{-\infty}^{\infty} \frac{dq}{2 \pi}
\frac{\cos( i q)}{4 J^2 s^2 q^2 + B^2} \nonumber \\
& =&  \frac{B}{4Js} - (-1)^i \exp\left(-i/\xi_B\right)
\label{finitefield}
\end{eqnarray}
without any adjustable parameters ($i \le N/2$). Here, we introduced the quantity
\begin{equation}
\label{xih}
\xi_B = \frac{2 J s}{B},
\end{equation}
which defines a characteristic length in units of the lattice spacing. The local magnetization decays exponentially
into the chain with a length scale $\xi_B$, which depends on the field $B$. Interestingly, the prefactor of the
alternating part is unity and independent of $J$ and $B$. The field therefore does not determine the strength of the
alternating response but only its range $\xi_B$. In this case, the edge effect is also large but not of order $N$, and
the thermodynamic contribution can be extracted using finite-size scaling, in contrast to Eq.~\eqref{local-mag}. The
local magnetization of a $N=100$ chain at intermediate field is shown in Fig.~\ref{fig:local-mag}, and good
quantitative agreement is found with the exact classical result [for very short chains Eq.~\eqref{finitefield} holds
only qualitatively, {see below}]. In first order in $(NB)^{-1}$ the total magnetization is obtained as $S = NB/(4J)$,
and $\xi_B$ thus varies with $S$ as $\xi_B \propto S^{-1}$. Since $S \le S_f$ (magnetic field $\le$ saturation field)
the limit $\xi_B > 1/2$ is implied. For relatively large $S$, when $\xi_B \ll N$, the alternating part becomes located
near the edges, and the local magnetization becomes essentially homogeneous in the interior of the chain, as it is also
observed qualitatively in the spin configurations shown in Fig.~\ref{fig:classicalconfigurations} for $S_n^2 = 0.3$.
The slope is determined as $g(S) = 1$, as in the $H_{AB}$ model, and consistent with the exact ferromagnetic slope
$g_F$ which is approached for $N \to \infty$. This finding may serve also as a measure of the quantitative accuracy of
Eq.~\eqref{finitefield}.

Eqs.~\eqref{local-mag} and \eqref{finitefield} describe two completely different physical regimes, and combined
they provide a qualitative description of the crossover. At small fields $B \ll Js/N$ the characteristic length $\xi_B$
exceeds the chain length and there is full interference between the edges, while at large fields $B \gg Js/N$ the
length $\xi_B$ is much shorter than $N$ and the two edges act independently. According to Fig.~\ref{fig:n6n7classical}
the crossover occurs when $S \approx 2 s$, which translates into $N \approx 4 \xi_B$ [which also implies $N \gg 2$ for
Eq.~\eqref{finitefield} to describe quantitatively the large-field regime, which is not always fulfilled in very short
chains].

In the thermodynamic limit $N\to \infty$ at very low fields $B\to 0$ both Eqs.~\eqref{local-mag} and
\eqref{finitefield} appear to be valid, but give contradictory results. This discrepancy is resolved if one takes care
of the order of limits. If $NB \to 0$ only Eq.~\eqref{local-mag} is applicable while Eq.~\eqref{finitefield} only holds
if $NB \to \infty$. Therefore, the thermodynamic limit $N \to \infty$ and the zero-field limit $B \to 0$ do not commute
in the classical model with edges, which has also been observed  for impurity effects in higher
dimensions.\cite{Eggert2007,Anfuso2006,Vojta2012}

In this section it has been shown that the difference in the $L$-band behavior between even and odd chains is captured
by the classical AFHC model. The analytic calculations for the spin densities naturally suggest a crossover in even
chains at the onset of interference between the edges, which the numerical results show to occur at $S \approx 2 s$ or
$N \approx 4 \xi_B$. Below this magnetization the edge spins of even chains can be magnetized with a low energy cost,
leading to a reduction of $g(S)$ predicted in Eq.~\eqref{smallg}. In odd chains on the other hand the ferrimagnetic
configuration at small $S$ prevents an easy magnetization of the edge spins, leading to a larger slope $g(S)$ in the
numerical results. It should be noted that the edge spins here are classical and are hence not related to the quantum
edge spins of the NLSM (Sec.~\ref{subsec:edgespins}). At magnetization above $2s$ the alternating spin density localizes
at the edges, and the distinction between even and odd chains largely disappears.

%%%%%%%%%%%%%%%%%%%%%%%%%%%%%%%%%%%%%%%%%%%%%%%%%%%%%%%%%%%%%%%%%%
% Spin Density
%%%%%%%%%%%%%%%%%%%%%%%%%%%%%%%%%%%%%%%%%%%%%%%%%%%%%%%%%%%%%%%%%%
\section{\label{sec:spindensity} Spin Density and Correlation Functions}

After having shown in the previous section that the even-odd effect can be rationalized with the help of the classical
spin densities and correlation functions, these quantities will be briefly examined for the quantum AFHC in the
antiferromagnetic region. The spin densities and correlation functions have to be symmetric under the parity operation
with respect to the center of the chain. This leads to an obvious difference in the spin density or wavefunctions
between even and odd chains. For even chains the parity operator interchanges the two sublattices, flipping the spins
of each sublattice. The symmetry competes with the antiferromagnetic order and leads to having the same nearest
neighbor $\langle s^z_i \rangle$ around the center, which is very small in magnitude. For odd $N$ there is no such
restriction.

\begin{figure}
\includegraphics[width=8.5cm]{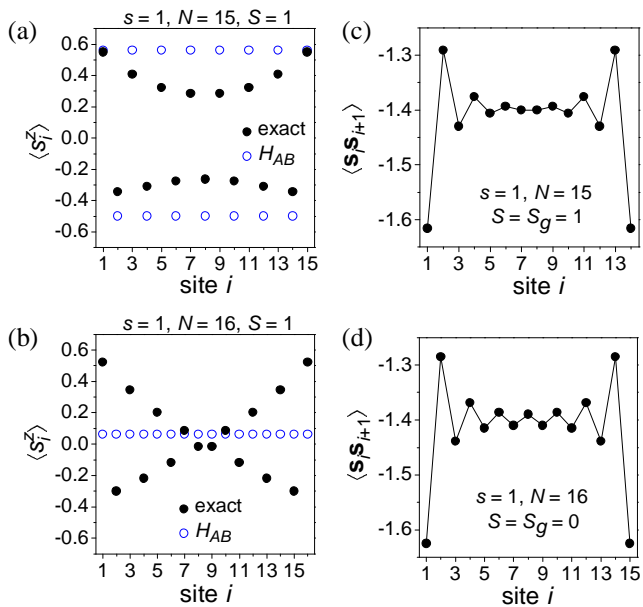}
\caption{\label{fig:spindensitiescorrelations} (Color online)
Spin density $\langle s_i^z \rangle$ and nearest-neighbor correlations $\langle  \mathbf{s}_i \cdot  \mathbf{s}_{i+1}
\rangle$ of the $s=1$ AFHC (black solid circles). (a) Spin density in the lowest $S = 1$ multiplet of the $N = 16$ even
chain and (b) for the $S = 1$ ground state of the $N = 15$ odd chain. (c) Correlations in the $S = 0$ ground state for
$N = 16$ and (d) in the $S = 1$ ground state for $N = 15$. For comparison in panels (a) and (b) the spin density of the
corresponding $H_{AB}$ model is shown (blue open circles). The nearest-neighbor correlations in the $H_{AB}$ model are
equal to $-1.125$ (not shown). Lines are guides to the eye.}
\end{figure}

A specific example is shown in Fig.~\ref{fig:spindensitiescorrelations}. The spin density of the $s=1$ chains with $N$
= 15 and 16 is plotted for the $S = 1$ lowest state (which is the ground and the first excited state respectively),
showcasing the differences for even and odd $N$ (note that for the ground state of the $N=16$ chain $\langle s^z_i
\rangle=0$ for all spins). The spin density is weaker at the center, while it increases approximately linearly going
towards the edges, where spins are less bound. The predictions of the $H_{AB}$ model are also plotted in the two
figures, and they miss the main features, even though they exhibit a difference between even and odd chains. For the
odd chain the $H_{AB}$ prediction clearly shows the antiferromagnetic order, while for the even chain it is uniform
(and hence small in magnitude).

For comparison, we shortly comment on the situation in long AFHCs. In Ref.~[\onlinecite{Sorensen1994}] the spin density
in the lowest $S=1$ state was numerically calculated with the DMRG method for $s=1$, $N=100$. A very good agreement has
been found with an exponential decay away from the chain ends, resulting from $s_{edge}=1/2$ edge spins. Additionally,
data for higher spin states $S>1$ confirmed the analytic picture of edge states and dilute boson-like bulk magnons in
long chains.\cite{Sorensen1993} However, for the $s=1$ chain the correlation length is about 6 sites, and the end spin
wavefunctions protrude accordingly from each end of the chain. The data shown here are for $N=15$ and 16 which is about
two times the correlation length, and the effective description by independent quantum edge spins is apparently not
appropriate.

Looking at the nearest-neighbor correlation functions for the ground states of the $N$=15 and 16 chains
[Fig.~\ref{fig:spindensitiescorrelations}(c),(d)], only weak differences between even and odd chains are observed.
Correlations are maximal at the edges, where the relatively loosely bound spins have more freedom to minimize
correlations in their vicinity. The strength of the nearest-neighbor correlations decreases towards the center. The
difference between the even and the odd chain is seen in the central region, where the correlation oscillates in
strength with position for $N=16$, similarly to what happens for the spins further out
[Fig.~\ref{fig:spindensitiescorrelations}(d)]. In contrast, for $N=15$ the strength of the central correlations does
not oscillate much with position [Fig.~\ref{fig:spindensitiescorrelations}(c)]. The $H_{AB}$ model predicts for both
cases uniform nearest-neighbor correlations equal to $-1.125$ and completely misses the central features.

\section{\label{sec:vbs} Valence Bond States}

The $H_{AB}$ model can describe the physics of the AFHC when N\'eel-type correlations prevail. Its main deficiency is
that it doesn't account for spatial fluctuations, leading to an infinite correlation length in this model and small
nearest neighbor quantum entanglement. The valence bond solid (VBS) is a complementary description, where strong
(singlet) entanglement between nearest neighbors is built in, and correlations are exponentially
decaying.\cite{AKLT,AKLT2,Arovas88} In contrast to the $H_{AB}$ model the VBS states also explicitly contain spin
degrees of freedom near the edges, and might hence better approximate the spatial fluctuations relevant for the
even-odd effect. We therefore compare and combine quantum VBS states with N\'eel-type $H_{AB}$ states in order to
understand better which effect plays a more dominant role.

\begin{figure}
\includegraphics[width=7cm]{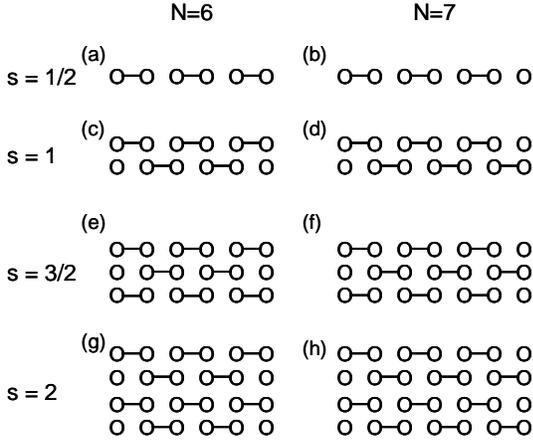}
\caption{\label{fig:vbs}
Sketches of valence bond states for $N=6$ and $7$ chains. Each column represents a spin $s$, each circle a spin $1/2$,
and each line a singlet bond. (a) Even and (b) odd completely dimerized $s=1/2$ chain. (c) Even and (d) odd VBS state
in an $s=1$ chain. (e) Even and (f) odd partially dimerized $s=3/2$ chain. (g) Even and (h) odd  VBS state in an $s=2$
chain.}
\end{figure}

\subsection{\label{sec:vbsconstr} Construction of Valence Bond States}

VBS states were originally introduced as translationally invariant ground states of exactly solvable integer spin
models with an excitation gap,\cite{AKLT, AKLT2} rigorously exemplifying the Haldane phase\cite{Haldane,Haldane2}  for
the first time. In general, valence bond states are formed by replacing the spin $s$ operators by symmetrized $2s$
spin-1/2 objects on each site, and then coupling pairs of spin-1/2 objects on different sites to form
singlets.\cite{Auerbach98} Many different valence bond states can   be constructed for a particular system in this way
depending on the coupling scheme of the spins, forming an overcomplete basis of the Hilbert space in the singlet
sector.\cite{Beach2006} As a simplification, we here restrict the valence bonds to connect nearest-neighbors evenly to
the right and the left as shown in Fig.~\ref{fig:vbs}. Because the spin-1/2 objects are symmetrized at each site, the
resulting correlations remain extended over a correlation length of several spins. For integer spin chains this
construction leads to a unique, and in case of periodic boundary conditions, translationally invariant state called a
VBS.\cite{AKLT,AKLT2} In case of half-integer $s$ the number of singlets between nearest neighbor sites of  the valence
bond wavefunction is different for two successive pairs, see  Fig.~\ref{fig:vbs}(e,f), therefore translational
invariance is lost. The depicted  state for each half-integer $s$ case is complemented by a state where all bonds are
shifted one lattice spacing to the right.  More generally, a half-integer spin VBS can be regarded as an integer spin
VBS with an additional spin-1/2 chain.  However, according to the magnetization profile in Eq.~(\ref{localsz}) the
residual free spin for odd $N$ spin-1/2 chains is located mostly in the center of the chain i.e. not as depicted in
Fig.~\ref{fig:vbs}(b,f) near the edge. The approximate ground state for half-integer $s$ is therefore formed by all
states where the residual free spin is delocalized.  However, using a suitable parent Hamiltonian based on projection
operators a unique trial VBS state can be defined as will be shown below.

Using the original idea from Affleck, Kennedy, Lieb and Tasaki, parent Hamiltonians with nearest neighbor VBS
wavefunction as the ground state can be constructed for integer $s$ by projecting out all parts where neighboring spins
couple to a total spin less than $s$. In case of open boundaries, this results exactly in the ground states shown in
Fig.~\ref{fig:vbs}(c,d,g,h) with $s$ unpaired spin $1/2$ objects at each end. These edge spins can form total spin
multiplets ranging from 0 to $s$, thus the ground state is $(s+1)^2$ times degenerate (including the degeneracy with
respect to $S^z$). Parent Hamiltonians can also be constructed for half-integer spin chains\cite{spin32dimer}, as will
be shown below for the case $s=3/2$.

In order to define parent Hamiltonians with exact VBS ground states it is useful to define (non-normalized) projection
operators acting on sites $i$ and $i+1$, which project out all states with total spin $(\mathbf{s}_i +
\mathbf{s}_{i+1})^2 < F$:
\begin{equation}
P_{i,i+1}^F=\frac{1}{K_F} \prod_{f=0}^{F-1} \left[(\mathbf{s}_i + \mathbf{s}_{i+1})^2-f(f+1)\right].
\label{eq:project}
\end{equation}
The constant $K_F>0$ is conveniently fixed such that the prefactor of $\mathbf{s}_i\cdot\mathbf{s}_{i+1}$ equals
$1$. The bond projection operators can be easily expressed as polynomials  of $\mathbf{s}_i \cdot
\mathbf{s}_{i+1}$.\cite{Auerbach98} The parent Hamiltonians for $s=1$ and 2 are then
\cite{AKLT,AKLT2,Auerbach98}
\begin{eqnarray}
H_{s=1} &=&\sum_{i} P_{i,i+1}^2 \nonumber \\
&=& \sum_i \left[ \mathbf{s}_i \cdot \mathbf{s}_{i+1} + \frac{1}{3}( \mathbf{s}_i \cdot \mathbf{s}_{i+1})^2+ \frac{2}{3} \right],\label{eq:vbs1} \\
H_{s=2} &=& \sum_{i} P_{i,i+1}^3 %\nonumber  \\
= \sum_i \left[ \mathbf{s}_i \cdot \mathbf{s}_{i+1} + \frac{2}{9}( \mathbf{s}_i \cdot \mathbf{s}_{i+1})^2 \right. \nonumber \\
 && + \left. \frac{1}{63}( \mathbf{s}_i \cdot \mathbf{s}_{i+1})^3+ \frac{10}{7} \right].\label{eq:vbs2}
\end{eqnarray}

For half-integer spin the situation is more complicated, because the total spin of two neighboring spins is
obviously alternating in the VBS states in Fig.~\ref{fig:vbs}(a,b,e,f) and cannot be fixed to a constant. One
solution is to use alternating parent Hamiltonians. In particular for $s=3/2$ one can choose
\begin{eqnarray}\label{eq:h1221}
H^{12}_{s=3/2}&=&\sum_{i} (P_{2i,2i+1}^2+ P_{2i+1,2i+2}^3),\nonumber\\
H^{21}_{s=3/2}&=&\sum_{i} (P_{2i,2i+1}^3+ P_{2i+1,2i+2}^2),
\end{eqnarray}
with two different ground state wavefunctions: $|\Psi_{12}\rangle$ with one singlet bond between the first two
sites and $|\Psi_{21}\rangle$ with two singlet bonds between the first two sites. For even $N$ only
$|\Psi_{21}\rangle$ is a reasonable VBS trial state, while for odd $N$ both states are equivalent, so that a
parity symmetric combination of the two must be formed.

In the following we analyze the overlaps and expectation values of the corresponding VBS states
$|\Psi_{VBS}\rangle$ depicted in Fig.~\ref{fig:vbs}(c)-(h). The states $|\Psi_{VBS}\rangle$ can be numerically
calculated as ground states of the parent Hamiltonians using the iterative power method and a projection onto
the $S$ subspace of interest or alternatively using an iterative method described in Appendix~\ref{sec:app}.

%%%%%%%%%%%%%%%%%%%%%%%%%%%%%%%%%%%%%%%%%%%%%%%%%%%%%%%%%%%%%%%%%%
% Comparison HAB vs VBS
%%%%%%%%%%%%%%%%%%%%%%%%%%%%%%%%%%%%%%%%%%%%%%%%%%%%%%%%%%%%%%%%%%
\subsection{\label{sec:vbshab} Comparison of the $H_{AB}$ Model with the VBS Model}

\begin{figure}
\includegraphics[width=7cm]{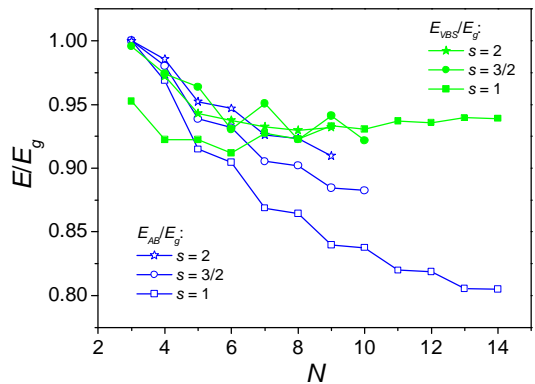}
\caption{\label{fig:energyoverlap}(Color online)
Variational energies $E_{VBS}$ and $E_{AB}$ of the AFHC Hamiltonian for the VBS states sketched in Fig.~\ref{fig:vbs}
(c)-(h) (green solid symbols)  and the ground states of the $H_{AB}$ model (blue open symbols). The energies are shown
with respect to the exact ground state energy $E_g$ of the AFHC for $s=1$ (squares), $s=3/2$ (circles), and $s=2$
(stars), as  a function of chain length $N$. Lines are guides to the eye.}
\end{figure}

\begin{figure*}
\includegraphics[width=17.5cm]{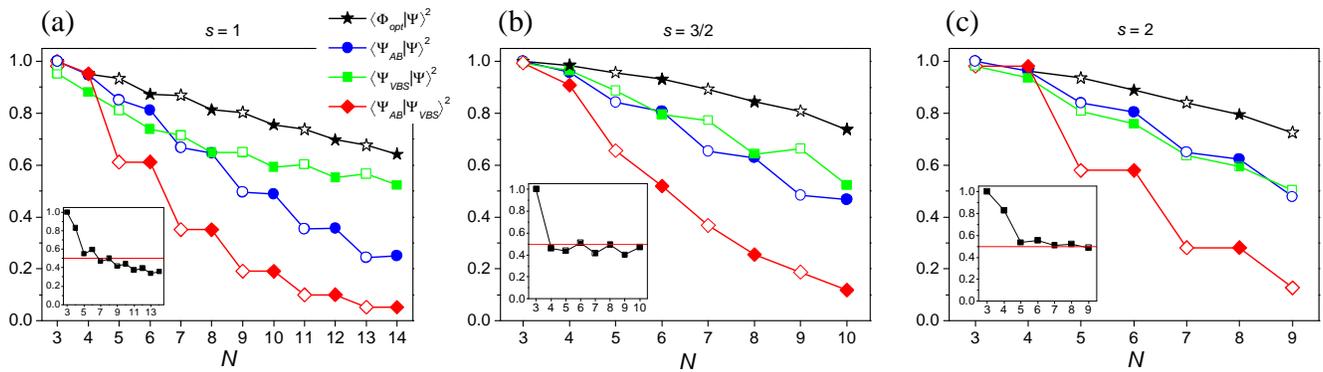}
\caption{\label{fig:wavefunctionoverlap}(Color online)
Squared overlap of the ground state wavefunction of the AFHC
$|\Psi\rangle$ with the VBS state ($\langle\Psi_{VBS}|\Psi\rangle^2$, green squares), with the $H_{AB}$ ground state
($\langle\Psi_{AB}|\Psi\rangle^2$, blue circles), and with an optimized linear combination
($|\Phi_{opt}\rangle=a|\Psi_{AB}\rangle +b|\Psi_{VBS}\rangle$, black stars) for the spin values (a) $s$=1, (b) $s$=3/2,
and (c) $s$=2 as function of chain length $N$. The relative overlap $\langle\Psi_{VBS}|\Psi_{AB}\rangle^2$ is shown for
comparison (red diamonds). In the insets $a/(a+b)$ is plotted (black squares). The red line corresponds to a value of
0.5. Lines are guides to the eye.}
\end{figure*}

In Fig.~\ref{fig:energyoverlap} the ratio of the variational energies $E_{AB}=\langle \Psi_{AB} |H|\Psi_{AB}\rangle$
and $E_{VBS}=\langle \Psi_{VBS} |H|\Psi_{VBS}\rangle$ over the exact ground state energy of the AFHC of Eq.
(\ref{eq:Hchain}) are plotted for $s=1$, 3/2, and 2. $|\Psi_{AB}\rangle$ is the ground state of the $H_{AB}$
Hamiltonian of Eq.~(\ref{eq:hab}), and $|\Psi_{VBS}\rangle$ is the ground state of the corresponding VBS parent
Hamiltonians of Eqs.~(\ref{eq:vbs1}), (\ref{eq:vbs2}), and (\ref{eq:h1221}). The accuracy of the variational energy of
the ground state of the $H_{AB}$ model drops off quickly with $N$, however for very small $N$ the $H_{AB}$ variational
energy is better than the VBS variational energy. Increasing $s$ also improves the quality of the $H_{AB}$ variational
energies, which agrees with the expectation that the $H_{AB}$ model is best suited for small $N$ and large $s$.  It
should be mentioned here that in the variational energy the even-odd effect appears to be reversed: ground state
energies for even $N$ are on the average slightly better approximated than for odd $N$ chains by $H_{AB}$, which is
opposite to what would be expected from the behavior of the excited states as described in Sec.~\ref{sec:evenodd}. For
the VBS energy ratios the variational energy for $s=1$ starts out relatively poor, but then improves with increasing
$N$. In contrast to the $H_{AB}$ model variational energies, the energy is generally estimated well also for large $N$
by $|\Psi_{VBS}\rangle$.

In Fig.~\ref{fig:wavefunctionoverlap} the overlaps of $|\Psi_{AB}\rangle$ and $|\Psi_{VBS}\rangle$ with the
ground state of the AFHC of Eq. (\ref{eq:Hchain}) are shown. These overlaps largely confirm the picture
discussed in the previous paragraph. For small $N$ and especially larger $s$ the $H_{AB}$ model has a slight
advantage over the VBS ground state, but then its overlap drops off quickly with $N$. Again the $H_{AB}$
overlaps are slightly better for even $N$ than for odd $N$. The overlap of $|\Psi_{VBS}\rangle$ on the other
hand is much less dependent on $s$ and also drops off slower with $N$. This shows that local quantum
entanglement is important for any $s$ and $N$, while the N\'eel-type order of $H_{AB}$ is relevant for large $s$
and small $N$. The overlap of the two ground states $|\Psi_{AB}\rangle$ and $|\Psi_{VBS}\rangle$ is also plotted
in Fig.~\ref{fig:wavefunctionoverlap}. Interestingly, both models give very similar wavefunctions up to $N=4$,
as can be concluded from the large overlap values.

To improve on the quality of the variational approximation both the $H_{AB}$ and VBS models were simultaneously
taken into account, by forming a trial state as a linear combination of the two corresponding wavefunctions,
namely the optimal wavefunction $|\Phi_{opt}\rangle = a|\Psi_{AB}\rangle + b|\Psi_{VBS}\rangle$ (the notation
always implies normalization). Its overlap with the AFHC ground state is also plotted in
Fig.~\ref{fig:wavefunctionoverlap} for the optimal combination of the variational parameters $a$ and $b$.  The
overlap improves in comparison with the two individual wavefunctions but still decreases with $N$, with a weak
dependence on $s$. In the insets of Fig.~\ref{fig:wavefunctionoverlap} the optimal ratio $a/(a+b)$ is plotted.
The overlap decrease with $N$ shows the importance of the VBS state for longer chains. It generally
increases with $s$, and the $H_{AB}$ model gains more weight as the increase of $s$ makes the AFHC ground state
more N\'eel ordered and less entangled.

Comparing the approximation of the ground state energy (Fig.~\ref{fig:energyoverlap}) with the overlap of the
ground state wavefunction (Fig.~\ref{fig:wavefunctionoverlap}) for the VBS model, the former hardly worsens with
$N$, while the overlap of the wavefunctions decreases. This is due to the fact, that the overlap of the VBS
model wavefunction with the AFHC wavefunctions of other low lying energy levels is still significant. Hence the
VBS wavefunction mostly mixes with the low lying AFHC energy levels.

%%%%%%%%%%%%%%%%%%%%%%%%%%%%%%%%%%%%%%%%%%%%%%%%%%%%%%%%%%%%%%%%%%
% Conclusions
%%%%%%%%%%%%%%%%%%%%%%%%%%%%%%%%%%%%%%%%%%%%%%%%%%%%%%%%%%%%%%%%%%
\section{\label{sec:concl} Summary and Conclusions}

In this work the structure of the lowest $S$ excitations in AFHCs of relatively short length $N$ but relatively large
spin magnitude $s$ has extensively been studied by contrasting the results of a broad array of theoretical tools and
approaches. The results of this paper are of relevance from at least three perspectives.

\textbf{Quantum vs spatial fluctuations}. First of all, the findings further our understanding of the physics in the
AFHC model. It has been demonstrated that there is a distinctive even-odd effect in the dependence of the lowest
energies $E(S)$ in each total-spin sector on $S$ or the $L$ band in short chains. The effect is markedly different to
the established even-odd effect in long chains, which is well understood in terms of the quantum edge-spin picture; the
arguments were given in Sec.~\ref{subsec:edgespins}. The different physics found in these two regimes justifies a
distinction into short and long chains, which represents a major finding of this work. The described even-odd effect
manifests itself in the antiferromagnetic region of the $L$-band spectrum (low $S$), but not in the ferromagnetic part
(high $S$). In the antiferromagnetic region even-odd effects can also be noticed e.g. in the ground-state spin, the
spin density, and the nearest-neighbor correlation functions. These are however straightforwardly explained by the
different symmetry properties of even and odd chains. The even-odd effect focused on in this work in contrast is not as
trivially traced back to the different symmetry properties of even and odd chains.

To elucidate the physics giving rise to this effect, different models were investigated, and the AFHC model was firstly
compared to the $H_{AB}$ model. Phenomenologically, the $H_{AB}$ model appeared as a promising candidate since it
naturally produces an $E(S) \propto S(S+1)$ energy dependence and an $L \& E$-band structure, as approximately observed
in short AFHCs (Fig.~\ref{fig:spectra}). While for odd chains the $H_{AB}$ model describes the $L$ band surprisingly
well in the full range of $S$ values, with a slight renormalization of the slope $g(S)$ or the effective gap
$\Delta_{AB}$, it fails to do so for even chains. The deviation is most pronounced in the antiferromagnetic region for
$S \lesssim 2s$, which is the even-odd effect, but also in the ferromagnetic region the deviation is significant. For
even-membered antiferromagnetic Heisenberg rings the energies predicted by the $H_{AB}$ model were previously shown to
become more accurate the larger $s$, and the $H_{AB}$ model was hence considered (semi-)classical in
nature.\cite{HabRing} Our results on the AFHC correct this view and point to the fact that the main characteristics of
the $H_{AB}$ model is the neglect of spatial fluctuations or implicit assumption of an infinite correlation length,
which consistently explains our findings. For instance, the spin densities are more homogeneous in odd than even chains
suggesting a better accuracy of the $H_{AB}$ predictions in the odd chains. Also, that the $H_{AB}$ model reproduces
energies and transition matrix elements extremely well for rings is now expected from the fact that in rings the
$L$-band states exhibit homogeneous spin densities by symmetry. These trends for rings, odd and even chains also
manifest themselves in the ferromagnetic region, as characterized by the slope $g_F$. For rings one finds $g_F = 1$,
which coincides with the prediction of the $H_{AB}$ model ($g=1$), while for chains (with $N>3$) $g_F
> 1$ holds, with the larger discrepancy for the even chains.

The predictions of the O(3) NLSM for short AFHCs were also analyzed. Interestingly, with neglected spatial fluctuations
the $H_{AB}$ model is reproduced, which underpins the role of spatial fluctuations and establishes the theoretical
basis of the $H_{AB}$ model. More importantly, the analysis demonstrated that the even-odd effect is not easily
reconciled within the NLSM. It in fact showed that the usual assumption of describing N\'eel order and uniform canting
as separate, weakly coupled degrees of freedom, which is exploited in or is even at the heart of many theories such as
bosonization, hydrodynamic theories, or those based on the NLSM, fails in short chains. In particular, it demonstrated
that the even-odd effect is distinct from the even-odd effects due to the quantum edge-spin model established for long
chains. The latter was furthered by an analysis of the VBS wavefunctions for AFHC systems. As e.g. demonstrated by the
analysis of the spin densities and correlations, differences between half-integer and integer spin chains are small for
short chains. The VBS results thus suggest that in short chains (with $s > 1/2$) the integer-spin VBS part in the total
VBS wavefunction is more relevant to the physics than the additional half-integer spin VBS part present in half-integer
spin chains. Somewhat surprisingly it was found that the VBS and $H_{AB}$ wavefunctions approximate the exact
wavefunctions nearly equally well (or equally poor) for small chains with relatively large spin magnitudes $s$. The
$H_{AB}$ and the quantum VBS models capture different aspects of the wavefunctions; each model has hence its strengths
and weaknesses with no clear advantage for one over the other.

Finally, the AFHC was also analyzed in the classical limit. The classical model describes the general trends and in
particular the even-odd effect very nicely, as shown by the numerical results, leaving little doubt that it captures
the essential physics. As main result it demonstrates the importance of spatial fluctuations in the even-odd effect.
Qualitatively, the even-odd effect can be related to the spatial inhomogeneities introduced by the spins at the edges
and their larger response to weak applied magnetic fields in the case of even chains [i.e. a smaller slope $g(S)$]. At
a quantitative level significant deviations remain unexplained, which reflects the fact that for the considered spin
magnitudes the classical limit is not yet reached and quantum fluctuations still play a significant role. As a
striking, yet so far unexplained consequence, the $L$ band of odd quantum chains is well described by the $H_{AB}$
model, which predicts the slope better than the classical model in Fig.~\ref{fig:n6n7classical}(b).

\textbf{Physical regimes in the AFHC model}. Having established fundamentally different behavior for short and long
chains, the question arises where the crossover between these regimes is located. It was first argued by Haldane that
in the thermodynamic limit $s$ determines the physical behavior, leading to a gap $\Delta_H \sim 0.4 J e^{-\pi s}$ in
the excitation spectrum for integer $s$, while half-integer spin chains have a linearly dispersing excitation
spectrum.\cite{Haldane,Haldane2,AffRev} In the framework of the renormalization group treatment of the NLSM in
Eq.~\eqref{nlsm} both integer and half-integer spin chains in fact show the same increase of the dimensionless coupling
constant $\gamma$ in the weak-coupling expansion.\cite{polyakov1975,fradkinbook} The length scale at which a
weak-coupling expansion breaks down is given by $e^{\pi s}$ irrespective of $s$ being integer or
half-integer.\cite{polyakov1975,fradkinbook} For integer spin chains this implies a gap proportional to the inverse
cut-off length $e^{-\pi s}$. For half-integer spin chains the topological term leads to a different physical behavior
which resembles that of the $s=1/2$ chain, i.e., a gapless critical behavior. While this difference is always observed
in the thermodynamic limit at small fields, it is important to realize that any relevant energy scale such as fields or
finite-size gaps will lead to a different renormalization flow. The physical behavior is then determined by the largest
energy scale or equivalently the smallest length scale.

In the case of finite chains there are several relevant length scales (energy scales), such as the chain length $N$ or
the correlation length due to finite fields $\xi_B$ in Eq.~\eqref{xih}. The length scale corresponding to the breakdown
of the weak-coupling expansion is
\begin{equation}
\label{Nc}
N_c = e^{\pi s},
\end{equation}
which corresponds to the correlation length in integer spin chains. Two fundamentally different physical regimes can be
identified (in zero field and temperature):

(1) $N_c \ll N$: This regime corresponds to the most studied case of the thermodynamic limit, where the famous
difference between integer and half-integer spin is observed. For finite chains with $N > N_c$ (long chains) it is
possible to clearly see the characteristic features of the thermodynamic limit by finite-size scaling (in the form of
characteristic corrections). The behavior can be well described by continuous quantum field theories; hence we call
this case the "renormalized continuous quantum regime".

(2) $N \ll N_c$: This regime of short chains was the main topic of this paper. The finite-size effects dominate and the
physical behavior is sensitive to the boundary condition and the geometry of the finite cluster, which leads to the
even-odd effect. It is fundamentally impossible to connect the unique behavior in this regime analytically to the
thermodynamic limit by finite-size scaling. Since many of the features are correctly reproduced by the corresponding
classical model but quantum effects are still important ({see below}), we call this case the "bounded quantum-classical
regime".

In finite magnetic fields the correlation length $\xi_B$ comes also into play. The above two regimes are present at low
fields, $N_c \ll \xi_B$ or $N \ll \xi_B$. The field gives rise however to a further regime where $\xi_B \ll N, N_c$,
which we call the "ferromagnetic regime". It is dominated by a relatively large magnetic field or large magnetization
$S \gg 2s, Ns/N_c$, and both short and long chains enter it under these conditions, obliterating the distinction
between the two low-field regimes. Our numerical results show that the correlations are dominated by the trend to align
all spins with the total spin. This behavior is continuously connected to the ferromagnetic region, and can be best
described by a hydrodynamic theory or by spin waves, which give analogous results.\cite{Eggert2007,Olav2011} The
behavior shows no fundamental difference between even and odd $N$ nor between integer and half-integer $s$. For
completeness it is mentioned that in addition there is also a finite temperature regime with smallest length scale
$J/T$, which is however not considered in this paper.

\begin{figure}
\includegraphics[width=7cm]{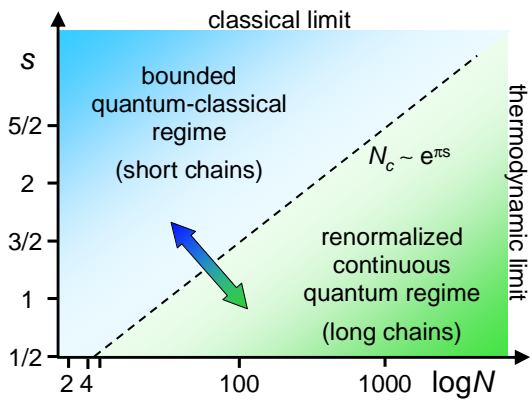}
\caption{\label{fig:NSdiagram}(Color online)
Sketch of the properties of the AFHC in the parameter space spanned by spin magnitude $s$ and chain length $N$ in a
$s$-$\log N$ plot. The region of the short chains is separated from the region of long chains by the characteristic
length $N_c(s) = e^{\pi s}$. The short-chain region connects to the classical limit $s \rightarrow \infty$ and the
long-chain region to the thermodynamical limit $N \rightarrow \infty$. The two regimes are suggested to exist also for
other, and potentially all antiferromagnetic Heisenberg clusters, and based on the characteristic properties they are
denoted as "bounded quantum-classical" and "renormalized continuous quantum" regime. }
\end{figure}

The above considerations demonstrate that the crossover from short to long is characterized by the length $N_c$, which
depends on $s$ and thus describes a boundary line as sketched in Fig.~\ref{fig:NSdiagram}. This plot was not completely
traced out by the numerical results (for obvious limitations in computational power), but the present work provides
strong pieces of evidence for its validity. For instance, since $N_c(1/2) \approx 5$, the region $N \ll N_c$ of short
chains is not reachable in $s=1/2$ chains, which is perfectly consistent with Fig.~\ref{fig:gAF}, where the variation
of $g_{AF}$ with $N$ is distinctly different for $s=1/2$, and the absence of the even-odd effect in the $s=1/2$ chain
as shown in Sec.~\ref{sec:spinhalf}. For $s=1$ the short-chain region starts to become available, but the short-chain
behavior may be realized only approximately, which is consistent with e.g. Fig.~\ref{fig:gAFgEvsN}. For $s \geq 3/2$,
however, the short-chain region is already available for significant chain lengths. This trend with $s$  is also
consistent with the notion that in the VBS picture the half-integer spin part in the VBS wavefunction (when present)
becomes less and less relevant the larger $s$ is, such that the physics is related to the integer spin part. Lastly,
the analysis of the classical chain model showed that for large $s$ the short-chain behavior is indeed present also in
chains with large, but finite $N$, as expected from Eq.~\eqref{Nc}. In particular, the slope $g(S)$ is suppressed for
large even $N$ in the low field limit, though only in a small range of fields $B \alt Js/N$. Generally, the short-chain
region becomes more accessible the larger $s$ and the smaller $N$ is, while the long-chain region is accessible for
relatively small $s$ and large $N$.

The region of small $s$ and large $N$ or long chains has been studied in great detail in the past and the physics can
be summarized as the renormalized continuous quantum regime. For short chains, in contrast, it was demonstrated in this
paper that many of the features, such as the spatial fluctuations, are qualitatively explained by the corresponding
classical model, and that in this sense short chains are classical. However, for experimentally accessible spin
magnitudes $s$ quantum fluctuations are clearly not negligible, as demonstrated e.g. by the slow convergence of the
quantum results to the classical limit or the significant overlap of the wavefunctions with the VBS states. A further
example is the superior performance of the $H_{AB}$ model over the classical model for the odd chains, and it remains
surprising that the slope $g(S)$ for odd chains follows the prediction of $H_{AB}$ well although the local correlations
do not. Short chains hence show both classical and quantum aspects, depending on the feature one is looking at, and in
this sense blur the distinction between classical and quantum physics. Hence we use the notation "bounded
quantum-classical regime" for this case. Remarkably, in this regime the physical behavior becomes largely independent
of $s$ and $N$, i.e., is generic.

\textbf{Implications for spin clusters in general}. At this point our results are of relevance also from a broader,
fundamental perspective. A diagram which at first sight is very similar to that in Fig.~\ref{fig:NSdiagram} was
proposed a decade ago based on studies on antiferromagnetic Heisenberg rings of relatively small size $N$ but with
relatively large spin magnitudes $s$.\cite{HabRing} It was in particular demonstrated that the $L\&E$-band picture and
the energies and matrix elements calculated with $H_{AB}$ become more accurate the larger $s$ and the smaller $N$
is.\cite{HabRing,Chiolero98} That is, the large $s$-small $N$ region was (erroneously) linked in this and subsequent
works to the validity of the $H_{AB}$ model, which lead to some inconsistencies.\cite{Review10} In the present work,
the situation is rectified by identifying the (classical) spatial fluctuations as the characterizing feature in this
regime. Both the "small" rings and short chains are apparently in the bounded quantum-classical regime. However, for
rings additionally the $H_{AB}$ model is an excellent approximation because of the symmetry-induced homogeneous spin
densities in their $L$-band states. For chains, in contrast, the $H_{AB}$ model is less appropriate.

The present work thus states more precisely the distinctive behavior of small rings anticipated in
Ref.~\onlinecite{HabRing} and puts it for the chain systems, through a very detailed numerical and theoretical
analysis, on a solid basis. The results lend credit to the idea that the two distinctive regimes are in fact generic
and present not only in rings and chains, but also in other and potentially many small antiferromagnetic Heisenberg
clusters, since one could generally expect that spatial fluctuations dominate over quantum fluctuations in these
systems. For the calculation of the quantum fluctuations introduced by magnetic anisotropy terms powerful theoretical
tools exist.\cite{guntmqt} In contrast, for the effects of the antiferromagnetic Heisenberg interactions in relatively
small lattices, a satisfying quantitative theory which takes into account the relevant effects, i.e., treats the
(classical) spatial fluctuations correctly and introduces the quantum effects, appears to be missing at the moment.
Developing it should be an attractive challenge for the future.

It is finally mentioned that clusters of a dozen exchange-coupled spin centers with relatively large spin magnitudes
$s$ are currently of high experimental relevance. An abundance of examples is provided by the class of molecular
nanomagnets,\cite{Gat06-book} which through synthetic chemistry has generated hundreds of magnetic molecules with
different arrangements of the metal centers. The number of spin centers ranges from 4 to a current maximum of 84, and
the spin magnitudes are typically $s=3/2$, 2, and 5/2 for transition metal clusters. Furthermore, the emerging field of
the artificially engineered spin clusters can be expected to provide many further attractive experimental
systems.\cite{Hirjibehedin06,Trimersurface,sufacerev} The key findings in this work should be of fundamental importance
to a variety of currently studied experimental systems, where mesoscopic effects are very important, and very promising
applications go hand in hand with interesting many-body effects.

\begin{acknowledgments}
The authors thank Ian Affleck and Frank Pollmann for many highly useful comments and fruitful discussions on the topic.
Partial funding by the Deutsche Forschungsgemeinschaft is thankfully acknowledged.
\end{acknowledgments}

\begin{appendix}

\section{\label{sec:base} Spin coupled basis}

In view of the SU(2) symmetry of Eq.~\eqref{eq:Hchain} it is convenient to perform numerical work directly in a basis
of eigenstates of the total spin operator $\mathbf{S}$. Since this is rarely done in the area of quantum spin systems,
some details shall be given here. The general procedure is given by Racah's methods and the irreducible tensor operator
(ITO) techniques.\cite{Bencini90}

In this paper the spin coupling scheme was used where at first the first two spins are coupled, and then successively the
next spin to the previous ones: $\mathbf{S}=(( \ldots((\mathbf{s}_1+\mathbf{s}_2)+\mathbf{s}_3)+\ldots)+\mathbf{s}_N)$.
This yields the SU(2) invariant basis states $|s_1 s_2 S_{12} s_3 S_{123} \ldots s_N S\rangle$ with intermediate spin
quantum numbers $S_{12}$, $S_{123}$, $\ldots$, $S_{1 \ldots N-1}$. This basis is exploited by expressing the
Hamiltonian in terms of ITOs.\cite{Bencini90}

The ITO $T^k(\mathbf{s}_i)$  of rank $k$ associated to the spin center $\mathbf{s}_i$ has  $2k+1$ components
$T_q^k(\mathbf{s}_i)$ with $q=-k,-k+1,\dots,k$. Coupling ITOs of different rank and different spins is generally
achieved through
\begin{eqnarray}
T_q^k(\{k_i\},\{\tilde{k}_{j}\})&= &
[[\dots[[T^{k_1}(\mathbf{s}_1) \times T^{k_2}(\mathbf{s}_2)]^{\tilde{k}_{2}} \times T^{k_3}(\mathbf{s}_3)]^{\tilde{k}_{3}} \nonumber \\  &&
\times \dots ]^{\tilde{k}_{N-1}}\times T^{k_N}(\mathbf{s}_N) ]_q^k,
\label{eq:TQK1}
\end{eqnarray}
where the  $\tilde{k}_{ j}\equiv k_{1\dots j}$  have to be populated according to the spin coupling scheme and the
intermediate spin quantum numbers. The tensor product of two ITOs thereby reads
\begin{equation}
[T^{k_i}(\mathbf{s}_i) \times T^{k_j}(\mathbf{s}_j)]_{Q}^{K}=\sum_{q_i,q_j}{\langle k_i k_j q_i q_j | K Q\rangle
T_{q_i}^{k_i}(\mathbf{s}_i) T_{q_j}^{k_j}(\mathbf{s}_j)}. \label{eq:TQK2}
\end{equation}
By repeated application of Eq.~\eqref{eq:TQK2} the coupling of the ITOs in Eq.~\eqref{eq:TQK1} can be recast into a sum
over the products of single-spin ITOs and Clebsch-Gordon coefficients $\langle k_i k_j q_i q_j | K Q\rangle$.

For pairwise interactions we introduce the notation $T^k_q(k_i k_j|\mathbf{s}_i \mathbf{s}_j)$, indicating a many-spin
ITO $T_q^k(\dots)$ in Eq.~(\ref{eq:TQK1}) with corresponding values  $k_i$ and $k_j$  for the single-spin ITOs
$T^{k_i}(\mathbf{s}_i)$ and $T^{k_j}(\mathbf{s}_j)$ and all $k_l=0$ if $l\neq i,j$. Note that the elementary ITOs are
defined as
\begin{eqnarray}
T^{1}_0(\mathbf{s}_i)=S^z_i,\quad T^{1}_{\pm 1}(\mathbf{s}_i)=\mp \frac{1}{\sqrt{2}}S^{\pm}_i,
\end{eqnarray}
while $T^{0}_0(\mathbf{s}_i)$ is the identity. For a Heisenberg system only the ITO representation of $\mathbf{s}_i
\cdot \mathbf{s}_j$ is needed, which is
\begin{equation}
   ( \mathbf{s}_i \cdot \mathbf{s}_j)=-\sqrt{3}   T_0^0(1 1|\mathbf{s}_i \mathbf{s}_j) .
\label{eq:}
\end{equation}

The parent Hamiltonians for the VBS wavefunctions in Eqs.~\eqref{eq:vbs1} and \eqref{eq:vbs2} include also higher-order
coupling terms $(\mathbf{s}_i \cdot \mathbf{s}_j)^n$, which lead to higher order polynomials of the ITOs
$T^{k_i}_{q_i}(\mathbf{s}_i)$ and $T^{k_j}_{q_j}(\mathbf{s}_j)$ respectively, e.g.~$(\mathbf{s}_i\cdot
\mathbf{s}_j)^2=\sum_{q_1,q_2}(-1)^{q_1+q_2}T_{q_1}^1(\mathbf{s}_i)T_{q_2}^1(\mathbf{s}_i)T_{-q_1}^1(\mathbf{s}_j)T_{-q_2}^1(\mathbf{s}_j)$.
These polynomials  can be successively reduced  by the building up principle \cite{bickerstaff1976}
\begin{equation}
T^{k_1}_{q_1}(\mathbf{s}_i) T^{k_2}_{q_2}(\mathbf{s}_i) = \sum_{kq}{\langle k_1 k_2 q_1 q_2|k q\rangle[T^{k_1}(\mathbf{s}_i) T^{k_2}(\mathbf{s}_i)]^k_q}
\label{eq:factor1}
\end{equation}
with
\begin{eqnarray}
\left[T^{k_1}(\mathbf{s}_i) T^{k_2}(\mathbf{s}_i)\right]^k_q&=&
(-1)^{2s+k} \sqrt{2k+1}
\left\{
\begin{array}{ccc}
k_1 & k_2 & k\cr
s & s & s\cr
\end{array}
\right\} \nonumber\\
&&\times
\frac{\langle s||T^{k_1}(\mathbf{s}_i)||s\rangle \langle s||T^{k_2}(\mathbf{s}_i)||s\rangle
}{\langle s||T^k(\mathbf{s}_i)||s\rangle} \nonumber \\
&&\times T^k_q(\mathbf{s}_i).
\label{eq:factor2}
\end{eqnarray}
The reduced matrix elements  are given by
\begin{eqnarray}
\langle s||T^{k}(\mathbf{s}_i)||s\rangle=k!\left(\frac{(2s+k+1)!}{2^k (2k)!(2s-k)!}\right)^{1/2}.
\end{eqnarray}
The biquadratic term then becomes\cite{borras1999}
\begin{eqnarray}
(\mathbf{s}_i\cdot \mathbf{s}_j)^2%&=&\sum_{q_1,q_2}(-1)^{q_1+q_2}T_{q_1}^1(\mathbf{s}_i)T_{q_2}^1(\mathbf{s}_i)T_{-q_1}^1(\mathbf{s}_j)T_{-q_2}^1(\mathbf{s}_j)\nonumber \\
%\\
&=&\sqrt{5}T_0^0(22|\mathbf{s}_i \mathbf{s}_j)+\frac{\sqrt{3}}{2}T_0^0(11|\mathbf{s}_i \mathbf{s}_j)+\frac{\mathbf{s}_i^2\cdot \mathbf{s}_j^2}{3}.\nonumber \\
\end{eqnarray}
For the $s=2$ VBS parent Hamiltonian also the $(\mathbf{s}_i\cdot \mathbf{s}_j)^3$ term is needed for which we obtain
\begin{eqnarray}
(\mathbf{s}_i\cdot \mathbf{s}_j)^3
&=&-\sqrt{7}T_0^0(33|\mathbf{s}_i \mathbf{s}_j)-2\sqrt{5}T_0^0(22|\mathbf{s}_i \mathbf{s}_j)\nonumber \\
&&-\frac{\sqrt{3}}{5}\left(3 \mathbf{s}_i^2\cdot \mathbf{s}_j^2-\mathbf{s}_i^2-\mathbf{s}_j^2+2\right)T_0^0(11|\mathbf{s}_i \mathbf{s}_j)\nonumber \\
&&-\frac{ \mathbf{s}_i^2\cdot \mathbf{s}_j^2}{6}.
\end{eqnarray}
Finally, the VBS parent Hamiltonians in terms of ITOs  read:
\begin{eqnarray}
%H_{s=1}& =& \frac{1}{3} \sum_i \left[\sqrt{5}T^0_0(22|\mathbf{s}_i\mathbf{s}_{i+1})
%-\frac{5\sqrt{3}}{2}T^0_0(11|\mathbf{s}_i\mathbf{s}_{i+1})\nonumber \right. \nonumber \\
%&& \left.+\frac{1}{3}\mathbf{s}_i^2 \mathbf{s}_{i+1}^2 +2 \right],\\
H_{s=1}& =& \frac{1}{3} \sum_i \left[\sqrt{5}T^0_0(22|\mathbf{s}_i\mathbf{s}_{i+1})
-\frac{5\sqrt{3}}{2}T^0_0(11|\mathbf{s}_i\mathbf{s}_{i+1})\nonumber \right. \nonumber \\
&& \left.+\frac{10}{3}\right],\\
H_{s=2}&=& \frac{1}{7} \sum_i\left[-\frac{\sqrt{7}}{9}T^0_0(33|\mathbf{s}_i\mathbf{s}_{i+1})
                        +\frac{4\sqrt{5}}{3}T^0_0(22|\mathbf{s}_i\mathbf{s}_{i+1})\right. \nonumber\\
&&\left. -\frac{42\sqrt{3}}{5} T^0_0(11|\mathbf{s}_i\mathbf{s}_{i+1})
                                 + 28 \right].
\end{eqnarray}

\section{\label{sec:app}Iterative construction of VBS wavefunctions}

VBS wavefunctions can be calculated in the spin coupled basis using iteration. For a dimer of two spins $s$ the VBS
wavefunction is known in the spin coupled basis. It is degenerate in $S$, and the maximal $S$ is $s$. The wavefunctions
are $|0\rangle_2=|s,s,0\rangle$ for total spin $S=0$ in an obvious notation, and go up to $|S \rangle_2 =|s,s,s\rangle$
for total spin $S=s$. If it is known what happens when a further spin is attached, i.e., if the VBS wavefunction for an
open chain of 3 spins with spin $s$ is known, then by iteration the VBS wavefunction for any $s$ chain of length $N$
can be calculated. In each iteration step the corresponding basis vectors are extended by one lattice site, i.e. from
$|S_{12},\ldots ,S_{1 \ldots N-1}\rangle_{N-1}$ to $|S_{12},\ldots ,S_{1\ldots N-1}, S\rangle_N$. The method finds the
unnormalized coefficients for a VBS wavefunction in the spin coupled basis, and the resulting wavefunction has
therefore to be normalized after completion of the iteration.

For an $s=1$ VBS chain [Figs.~\ref{fig:vbs}(c), (d)] the iteration reads
\begin{eqnarray}
|0\rangle_N&=&|1\rangle_{N-1},\\\nonumber
|1\rangle_N&=&|1\rangle_{N-1}- \frac{\sqrt{3}}{2}|0\rangle_{N-1}.
\label{eq:vbsalgo1}
\end{eqnarray}
The basis functions of $|1\rangle_{N}$ and $|0\rangle_{N}$ are orthogonal. The number of relevant basis functions for
the $s=1$ chain grows with $N$ like the Fibonacci numbers. For an $s=2$ VBS chain [Figs.~\ref{fig:vbs}(g), (h)] one
finds
\begin{eqnarray}
|0\rangle_N&=&|2\rangle_{N-1},\\\nonumber
|1\rangle_N&=&|2\rangle_{N-1}-\sqrt{\frac{5}{7}}|1\rangle_{N-1},\\\nonumber
|2\rangle_N&=&|2\rangle_{N-1}-\sqrt{\frac{135}{49}}|1\rangle_{N-1}+\sqrt{\frac{80}{49}}|0\rangle_{N-1},
\label{eq:vbsalgo2}
\end{eqnarray}
and for an $s=3$ VBS chain
\begin{eqnarray}
|0\rangle_N&=&|3\rangle_{N-1},\\\nonumber
|1\rangle_N&=&|3\rangle_{N-1}-\sqrt{\frac{7}{10}}|2\rangle_{N-1},\\\nonumber
|2\rangle_N&=&|3\rangle_{N-1}-\sqrt{\frac{7}{3}}|2\rangle_{N-1}+\sqrt{\frac{7}{6}}|1\rangle_{N-1},\\\nonumber
|3\rangle_N&=&|3\rangle_{N-1}-\sqrt{\frac{56}{10}}|2\rangle_{N-1}+\sqrt{\frac{28}{3}}|1\rangle_{N-1}\\\nonumber
  &&-\sqrt{\frac{175}{36}}|0\rangle_{N-1}.
\label{eq:vbsalgo3}
\end{eqnarray}

For the dimerized $s=3/2$ valence bond function [Figs.~\ref{fig:vbs}(e), (f)] with only one singlet bond between the
first two spins one obtains
\begin{eqnarray}
|0\rangle_{2N}&=&|3/2\rangle_{2N-1},\\\nonumber
|1\rangle_{2N}&=&|3/2\rangle_{2N-1}+\sqrt{\frac{2}{5}}|1/2\rangle_{2N-1},\\\nonumber
|2\rangle_{2N}&=&|3/2\rangle_{2N-1}+\sqrt{2}|1/2\rangle_{2N-1},\\\nonumber
|1/2\rangle_{2N-1}&=&|2\rangle_{2N-2}-\sqrt{\frac{1}{3}}|1\rangle_{2N-2},\\\nonumber
|3/2\rangle_{2N-1}&=&|2\rangle_{2N-2}-\sqrt{\frac{5}{3}}|1\rangle_{2N-2}+\sqrt{\frac{4}{5}}|0\rangle_{2N-2},
\label{eq:vbsalgo321}
\end{eqnarray}
and for the dimerized $s=3/2$ valence bond function with two singlet bonds between the first two spins holds
\begin{eqnarray}
|0\rangle_{2N}&=&|3/2\rangle_{2N-1},\\\nonumber
|1\rangle_{2N}&=&|3/2\rangle_{2N-1}-\sqrt{\frac{32}{25}}|1/2\rangle_{2N-1},\\\nonumber
|1/2\rangle_{2N-1}&=&|1\rangle_{2N-2},\\\nonumber
|3/2\rangle_{2N-1}&=&|1\rangle_{2N-2}-\sqrt{\frac{27}{25}}|0\rangle_{2N-2}.
\label{eq:vbsalgo322}
\end{eqnarray}

\section{Technical details in the derivation of the O(3) NSLM}\label{derivation_nlsm}

In this appendix the O(3) NSLM is derived in detail. With the decomposition (\ref{decomposition}) the action in
(\ref{action_manyspin}) is evaluated. First, expanding the term $J s^2 \sum_{i=1}^N \mathbf{\Omega}_i \cdot
\mathbf{\Omega}_{i+1}$ up to order $|\mathbf{l}/s|^2$:
\begin{eqnarray}
\mathbf{\Omega}_i\cdot\mathbf{\Omega}_{i+1}&\approx& -\mathbf{n}(x_i)\mathbf{n}(x_{i+1}) \left[1-\frac{\mathbf{l}^2(x_i)}{2s^2}-\frac{\mathbf{l}^2(x_{i+1})}{2s^2}\right]\nonumber\\
&&+\frac{\mathbf{l}(x_i)\mathbf{l}(x_{i+1})}{s^2}\\
& & +(-1)^{i+1}\left[\mathbf{n}(x_{i})\frac{\mathbf{l}(x_{i+1})}{s}-\mathbf{n}(x_{i+1})\frac{\mathbf{l}(x_{i})}{s}\right]\nonumber.
\end{eqnarray}
Differences of the N\'eel fields can be approximated by derivatives which allow to write
$\mathbf{n}(x_i)\mathbf{n}(x_{i+1})=1-\frac{1}{2}[\mathbf{n}(x_i)-\mathbf{n}(x_{i+1})]^2\approx
1-\frac{a^2}{2}[\partial_x \mathbf{n}(x_i)]^2$. Then up to a constant term:
\begin{eqnarray}
\mathbf{\Omega}_i\cdot\mathbf{\Omega}_{i+1}&\approx& \frac{a^2}{2}\left[\partial_x\mathbf{n}(x_i)\right]^2+\frac{\left(\mathbf{l}(x_i)+\mathbf{l}(x_{i+1})\right)^2}{2s^2}\\
&&+(-1)^{i+1}\left[\mathbf{n}(x_{i})\frac{\mathbf{l}(x_{i+1})}{s}-\mathbf{n}(x_{i+1})\frac{\mathbf{l}(x_{i})}{s}\right]\nonumber\\
&&+\mathcal{O}\left[a^2\frac{\mathbf{l}^2}{s^2}(\partial_x\mathbf{n})^2\right]\nonumber.
\end{eqnarray}
The alternating term requires a careful treatment of the boundary conditions. For periodic boundary conditions
and $N$ even the term can be neglected as can be seen by writing
$\mathbf{n}(x_{i})\frac{\mathbf{l}(x_{i+1})}{s}\approx-\frac{a}{s}\partial_x\mathbf{n}(x_{i+1})\mathbf{l}(x_{i+1})$
and $\mathbf{n}(x_{i+1})\frac{\mathbf{l}(x_{i})}{s}\approx\frac{a}{s}\partial_x\mathbf{n}(x_i)\mathbf{l}(x_i)$
which directly yields
\begin{equation}\label{boundaryterms0A}
Js^2\sum_{i=1}^{N}(-1)^{i+1}\left[\mathbf{n}(x_{i})\frac{\mathbf{l}(x_{i+1})}{s}-\mathbf{n}(x_{i+1})\frac{\mathbf{l}(x_{i})}{s}\right]
\approx 0.
\end{equation}
For open boundary conditions additional boundary terms remain. Discussion of these terms is left for later and
periodic boundary conditions are considered now. Introducing $L=Na$ and taking the continuum limit:
\begin{equation}\label{continuumlimit}
Js^2\sum_{i=1}^N\mathbf{\Omega}_i\cdot\mathbf{\Omega}_{i+1} \longrightarrow J\int_0^Ldx\left[\frac{a s^2}{2}\left(\frac{\partial \mathbf{n}}{\partial x}\right)^2+\frac{2}{a}\mathbf{l}^2\right].
\end{equation}
Next, the imaginary part in Eq.~\eqref{action_manyspin} is evaluated. The Berry phase is antisymmetric under
inversion $\omega[\mathbf{\Omega}_i]=-\omega[-\mathbf{\Omega}_i]$, hence
\begin{eqnarray}\label{imaginary_part}
i s \sum_{i=1}^N\omega[\mathbf{\Omega}_i]&=&-i s\sum_{i=1}^N \left\{ (-1)^{i+1}\omega[\mathbf{n}(x_i)] \right. \nonumber\\
&&-i\int_0^\beta d\tau\left[\mathbf{n}(x_i,\tau)\times \frac{\partial \mathbf{n}(x_i,\tau)}{\partial \tau}\right] \nonumber \\
&& \left. \cdot \mathbf{l}(x_i,\tau) \right\}.
\end{eqnarray}
In Sec.~\ref{sec:nlsm} the first term in Eq.~\eqref{imaginary_part} is of "topological" significance. The second
one in contrast enters in the classical equation of motion.

Up to an additive constant the total action reads
\begin{eqnarray}
A&=&\int_0^L dx\int_0^\beta d\tau \left[\frac{Jas^2}{2}\left(\frac{\partial\mathbf{n}}{\partial x}\right)^2+\frac{2J}{a}\mathbf{l}^2\right.\nonumber\\
&&\left.-\frac{i}{a} \left(\mathbf{n}\times \frac{\partial \mathbf{n}}{\partial \tau}\right)\cdot\mathbf{l}\right]+A_{\rm{top}}.
\end{eqnarray}
Completing the square the functional integration over $\mathbf{l}$ can be performed, giving
$\mathbf{l}=\frac{i}{4J}(\mathbf{n}\times \partial_\tau \mathbf{n})$. Thus, the field $\mathbf{l}$ generates rotations
on $\mathbf{n}$. Note that the constraint $\mathbf{l} \cdot \mathbf{n} = 0$ is automatically fulfilled. Overall
normalization constants are left out. Finally, the O(3) NSLM is generated, with a $\theta$ term as the effective action
for the Heisenberg chain in the large $s$ limit, see Eq.~\eqref{nlsm}.

Finally, Eq.~\eqref{boundaryterms0A} is considered for open boundaries. Using
$\mathbf{n}(x_{i})\frac{\mathbf{l}(x_{i+1})}{s}\approx-\frac{a}{s}\partial_x\mathbf{n}(x_{i+1})\mathbf{l}(x_{i+1})$
and $\mathbf{n}(x_{i+1})\frac{\mathbf{l}(x_{i})}{s}\approx\frac{a}{s}\partial_x\mathbf{n}(x_i)\mathbf{l}(x_i)$
the following boundary terms are obtained:
\begin{eqnarray}\label{boundaryterms2}
&&Js^2\sum_{i=1}^{N-1}(-1)^{i+1}\left[\mathbf{n}(x_{i})\frac{\mathbf{l}(x_{i+1})}{s}-\mathbf{n}(x_{i+1})\frac{\mathbf{l}(x_{i})}{s}\right]\nonumber \\
&&\approx\left\{\begin{array}{lll}
-a s J\left[\partial_x\mathbf{n}(0)\mathbf{l}(0)+\partial_x\mathbf{n}(L)\mathbf{l}(L)\right]\quad\mbox{$N$ even}& \quad \\ \\
-a s J\left[\partial_x\mathbf{n}(0)\mathbf{l}(0)-\partial_x\mathbf{n}(L)\mathbf{l}(L)\right]\quad\mbox{$N$ odd}&
\end{array}\right.
\end{eqnarray}
These terms have a scaling dimension of one order higher than the bulk terms in the action in Eq.~\eqref{nlsm}. Note
that they are of the same order as the higher order terms in the Euler-Maclaurin sum formula that occur when going from
the discrete sum to the continuum integral, see Eq.~\eqref{continuumlimit}. It is assumed that these terms are small.

\end{appendix}

%\bibliography{paperchain}
\bibliography{paperchain16Nov2012}

\end{document}